\documentclass[showpacs,twocolumn,showkeys,prd,aps,10pt,nofootinbib,final,floatfix]{revtex4-1}
\usepackage{amssymb}
\usepackage{graphicx,amsmath,color}
\usepackage{graphicx}
\usepackage{psfrag}
\usepackage{color}

\begin{document}

\title{Parameter estimation for inspiraling eccentric compact binaries\\
including pericenter precession}
\author{Bal\'{a}zs Mik\'{o}czi$^{1}$, Bence Kocsis$^{2}$, P\'{e}ter Forg\'{a}cs$^{1,3}$,
and M\'{a}ty\'{a}s Vas\'{u}th$^{1}$}

\affiliation{$^{1}$Research Institute for Particle and Nuclear
Physics, Wigner RCP H-1525 Budapest 114, P.O.~Box 49, Hungary}
\affiliation{$^{2}$ Harvard-Smithsonian Center for Astrophysics,60
Garden Street, Cambridge, Massachusetts 02138, USA}
\affiliation{$^3$LMPT, CNRS-UMR 6083, Universit\'e de Tours, Parc de
Grandmont, 37200 Tours, France}

\begin{abstract}
Inspiraling supermassive black hole binary systems with high orbital
eccentricity are important sources for space-based gravitational
wave (GW) observatories like the Laser Interferometer Space Antenna
(LISA). Eccentricity adds orbital harmonics to the Fourier transform
of the GW signal, and relativistic pericenter precession leads to a
three-way splitting of each harmonic peak. We study the parameter
estimation accuracy for such waveforms with different initial
eccentricity using the Fisher matrix method and a Monte Carlo
sampling of the initial binary orientation. The eccentricity
improves the parameter estimation by breaking degeneracies between
different parameters. In particular, we find that the source
localization precision improves significantly for higher-mass
binaries due to eccentricity. The typical sky position errors are
$\sim1\,$deg for a nonspinning, $10^7\,M_{\odot }$ equal-mass binary
at redshift $z=1$, if the initial eccentricity 1 yr before merger is
$e_0\sim 0.6$. Pericenter precession does not affect the source
localization accuracy significantly, but it does further improve the
mass and eccentricity estimation accuracy systematically by a factor
of 3--10 for masses between $10^6\,M_{\odot }$ and $10^7\,M_{\odot
}$ for $e_0 \sim 0.3$.
\end{abstract}

\maketitle

\section{Introduction}

The inspiral and merger of compact binary systems of black holes are
important sources of gravitational waves (GWs) for the proposed
space-based GW missions such as the Laser Interferometer Space
Antenna (LISA) \cite{LISA1} or the European New Gravitational Wave
Observatory (NGO/eLISA) \cite{eLISA1}. The detectable frequency band
for these instruments will be around $10^{-4}$ to $10^{-1}\,$Hz
\cite{LISA2} which corresponds to the inspiral of two
$(10^{4}-10^{7})M_{\odot }$ black holes. As the sources detected by
LISA/NGO will be loud with a large signal-to-noise ratio in general,
an ideal method for parameter extraction is matched filtering
\cite{Matched}.

An effective matched filtering requires an accurate model of the
emitted GWs. In this technique the detected signal output is
cross-correlated with theoretical waveform templates. In particular,
matched filtering is sensitive to the phase information of the
waveform, and a high correlation between the signal and template
allows one to make predictions on the source parameters
\cite{FC,CF}.

Many previous studies in the literature adopted waveforms generated
by binaries in circular orbits (see Refs.
\cite{Thorne,Cutler98,Vecchio,BBW,LH,Arun,TS,LH08,McWilliams,
HH,Bence-apj06,Bence-prd,Bence-apj08} for LISA parameter
estimation). This is due to the expectation that the orbit of the
binary will circularize due to the emission of GWs \cite{PM,P}.

Nevertheless, there are a number of reasons to expect that at least
some LISA sources may be eccentric. If the binary is embedded in a
gaseous disk, it can remain eccentric until the final year of the
inspiral \cite{AN,MacFadyen,Cuadra,Sesana}. The interaction of the
supermassive black hole (SMBH) binary with a population of stars
also increases its eccentricity \cite{Preto,Matsubayashi,Lockmann}.
The eccentricity can also be excited by the Kozai mechanism and
relativistic orbital resonances in hierarchial triples
\cite{Wen,HL,Seto2,AP,Naoz} or by a triaxial potential
\cite{HattoriYoshii,MerrittVasiliev}, and may be typical for extreme
mass ratio inspirals \cite{AmaroSeoane07,AmaroSeoane12}. Further,
black hole binaries in dense galactic nuclei formed by GW emission
during close encounters remain very eccentric until merger
\cite{Bence,KL}. Population synthesis and binary evolutionary models
show that a fraction of stellar compact object binaries may also be
eccentric for ground-based (Advanced LIGO/VIRGO and Einstein
Telescope) and space-based detectors (DECIGO) \cite{Kowalska}.

Including eccentricity in the waveform may be essential for the
detection of inspiraling eccentric binaries with matched filtering
and to avoid a systematic bias in the parameter estimation
\cite{CV}. Using circular templates to detect waveforms with
eccentricities $e_0\gtrsim 0.1$, leads to a significant loss of
signal-to-noise ratio for ground-based detectors such as LIGO and
VIRGO \cite{MP,BrownZimmer}. A similar conclusion was reached for
eccentric massive black hole binaries detected with LISA
\cite{PorSe}. The orbital evolution and waveforms have been
developed to first and second post-Newtonian (PN) order, including
spin-orbit and spin-spin contributions for eccentric
orbits~\cite{KMG,KleinJetzer,GopakumarSchaefer,TessmerSchaefer,Hinder,CornishKey}.

To assess the astrophysical impact of planned GW instruments, it is
essential to estimate the expected parameter measurement precision
of typical GW sources. This may be done by injecting a simulated GW
signal into synthetic detector noise and carrying out a Monte Carlo
Markov Chain (MCMC)-based matched filtering search for a
parametrized template model to recover the posterior distribution
function (PDF) of the estimated source parameters \cite{CP}. Porter
and Sesana \cite{PorSe} investigated the cases of low-mass
($100M_{\odot }$) and high ($10^{4}M_{\odot }$) mass black hole
binaries on eccentric orbits using nonspinning, restricted 2 PN
waveforms. They concluded that eccentricity can significantly bias
the recovered parameters of the source for LISA if circular
templates are used even if the eccentricity is as small as $e \sim
10^{-4}$. More recently, Key and Cornish \cite{KeyCornish} extended
that study by using an effective 1.5PN waveform for inspiraling
eccentric SMBHs [with $m\sim(10^{5}-10^{7})M_{\odot}$] taking into
account eccentricity and spin effects in the template model. They
found that the eccentricity measurement errors are of order $\Delta
e \sim 10^{-3}$ for a range of mass ratios and a particular choice
of angular parameters.

Since the parameter space is large, $17$-dimensional for an
eccentric spinning binary, state-of-the-art MCMC calculations are
numerically too expensive to explore the full range of source
parameters. However, for a large signal-to-noise ratio (SNR), the
PDF may be well approximated by an ellipsoid, and the parameter
measurement errors can be estimated very efficiently using the
Fisher matrix method \cite{FC,CV}. Using this method, it has been
shown that different source inclinations and sky locations lead to a
wide range of parameter measurement errors subtending many orders of
magnitude \cite{Hughes,Vecchio,BBW,LH,LH08}. In this study, we carry
out a Fisher matrix analysis to investigate the possible range of
parameter estimation errors for eccentric binaries.

Only a few studies have investigated the LISA parameter estimation
accuracy for eccentric inspiraling sources using the Fisher matrix
method (cf. Refs. \cite{Cutler98,Hughes,Vecchio,BBW,LH,LH08,Arun,TS}
for circular inspirals). Barack and Cutler \cite{BC} investigated
the LISA errors for highly eccentric stellar mass compact objects
inspiraling into a SMBH. They found that the influence of
eccentricities on $\Delta \mathcal{M}/\mathcal{M}\sim 10^{-4}$
(error of the chirp mass), $\Delta e_{0}\sim 10^{-4}$ (error of
initial eccentricity) and $\Delta \Omega_{S}\sim 10^{-4}$ (angular
resolution error) is not substantial; the error estimates do not
differ much from those obtained for circular orbits \cite{Cutler98}.
However, they assumed only an arbitrarily chosen, single set of
orientations, which may not be representative of the typical errors.
Yunes et al. \cite{Yunes} provided ready-to-use analytic expressions
for the Fourier waveforms of moderately eccentric sources. They have
shown that eccentricity increases the detectable mass range of GW
detectors toward higher masses by enhancing the orbital harmonics
\cite{Arun,TS}. Yagi and Tanaka \cite{Tanaka} investigated the LISA
errors for various alternative theories of gravity for spinning,
small-eccentricity inspiraling SMBH binaries ($e_0\sim 0.01$ at 1 yr
before merger), using restricted 2 PN waveforms, neglecting higher
orbital harmonics and apsidal precession in the waveform. They have
found that the eccentricity and the spin-orbit interaction reduce
the parameter errors by an order of magnitude for spinning SMBHs in
massive graviton theories, but not in Brans-Dicke-type theories.

Neither of the previous systematic Fisher matrix studies of parameter errors
included the effects of relativistic pericenter precession for eccentric
sources. However, precession effects introduce an additional feature in the
waveform, and have the potential to break the degeneracy between parameter
errors \cite{Bence-prd}. In particular, spin-orbit precession has been shown
to improve the source localization precision substantially during the last
day of the inspiral \cite{Vecchio,LH,LH08}. Similarly, GR pericenter
precession may also be expected to improve the LISA parameter measurement
accuracy. In fact, since pericenter precession enters at a lower PN order,
this improvement could take place well before the binary reaches merger.
Localizing the source before merger could be used to provide triggers for
electromagnetic (EM) facilities to search for the EM counterpart \cite%
{Bence-apj08}. A coincident GW and EM observation of the same source
could have far-reaching astrophysical implications
\cite{Schutz,HH,Bence-apj06,Bence-apj08}

In the present paper, we carry out a systematic parameter estimation
study for inspiraling SMBH binaries, taking into account both
orbital eccentricity and the relativistic pericenter precession
effect. We account for the evolution of the semimajor axis and
eccentricity in our waveforms to leading order due to GW emission
\cite{Whalquist,MP,Pierro,BC}, but we neglect higher-order PN
contributions and spin effects. We compute the waveform in the
frequency domain using the stationary phase approximation (SPA, see
Refs. \cite{Yunes,Tessmer,ecc1,ecc2,Seto}) and derive the
signal-to-noise ratio (SNR) and the Fisher information matrix using
a Fourier-Bessel analysis for the parameter estimation of eccentric
sources. To explore the possible range of parameter errors, we
generate a Monte Carlo sample of binaries with random orientations
and vary the masses and initial eccentricities systematically over a
wide range relevant for LISA. We calculate the parameter errors for
the standard three-arm LISA/NGO configuration, as well as for a
descoped detector configuration, where one of the two independent
interferometers is removed.

In Sec.~II, we summarize the basic formulas describing eccentric
waveforms in the leading quadrupole approximation, using a
Fourier-Bessel decomposition. In Sec.~III, we derive the frequency
domain waveforms and the LISA detector response. After a brief
introduction of parameter estimation using the Fisher matrix method
in Sec. IV, we present results for specific systems in Sec. V. We
summarize our conclusions in Sec VI. Some details of the
calculations are described in Appendixes A and B.

We use geometrical units $G=c=1$.

\section{Time-dependent eccentric waveforms}

\begin{figure}[th]
\caption{The geometry of an eccentric orbit. The coordinate system
$(x,y,z)$ is defined by the initial orbit, where the x axis points
in the direction of the pericenter and the z axis is parallel to the
orbital angular momentum vector. In the reduced Kepler problem the
body with mass $\protect\mu=m_{1}m_{2}/m$ is orbiting the central
mass $m=m_{1}+m_{2}$; the separation vector is
$r=a_{0}(1-e_{0}^{2})/(1+e_{0}\cos \protect\phi )$, where $e_{0}$ is
the orbital eccentricity; $a_{0}=m^{1/3}(2\protect\pi \protect\nu
_{0})^{2/3}$ (here $\protect\nu _{0}$ is the orbital frequency) is
the semimajor axis; $\protect\phi $ is the true anomaly (the angle
between pericenter and the separation vector); and $\protect\gamma
_{0}$ is the pericenter position. The Kepler equation determines the
evolution of the time parameter: $\protect\xi -e_{0}\sin \protect\xi
=2\protect\pi \protect\nu _{0}(t-t_{0})$, where $\protect\xi $ is
the eccentric anomaly [$\tan \protect\xi
/2=\protect\sqrt{(1-e_{0})/(1+e_{0})}\tan \protect\phi /2$]. The
adiabatic evolution of the eccentric orbit is driven by the
pericenter precession (1 PN effect) and the inspiral (2.5 PN effect)
of the compact binary due to gravitational radiation. }
\label{geometry}
\begin{center}
\includegraphics[width=0.42\textwidth]{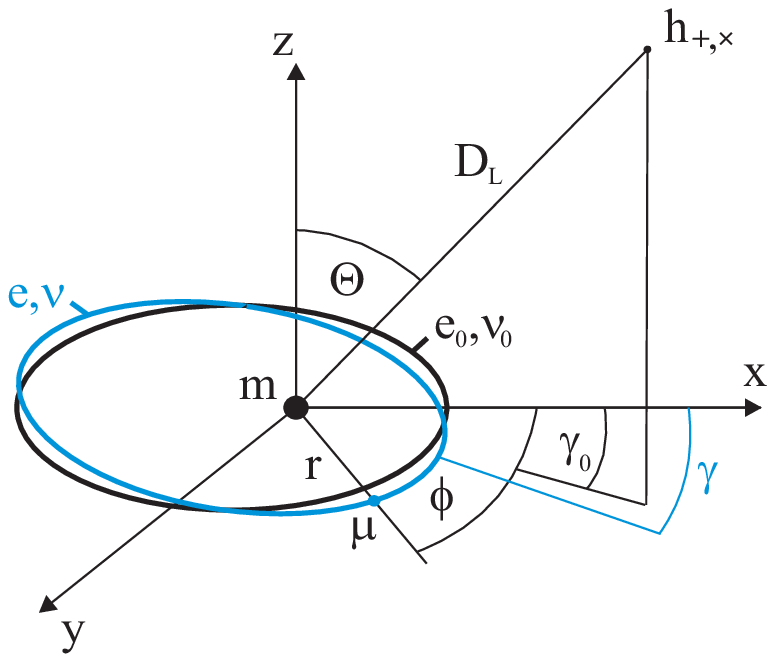}
\end{center}
\end{figure}

To leading order, the waveform emitted by a binary moving on a
Keplerian orbit can be computed by the quadrupole approximation. In
this approach the observer (i.e. the interferometric detector) is
assumed to be far from the source and higher-order contributions;
e.g., the effects of the spins and higher multipole moments are
neglected, but the orbit is corrected for the effect of
\textit{pericenter precession}. For such precessing Keplerian
orbits, the eccentric waveforms are given in Ref.~\cite{Whalquist}.
We have rewritten the leading-order quadrupole tensor and
transformed to the \textit{transverse-traceless gauge}, which gives
\begin{widetext}
\begin{eqnarray}
h_{\times }(\phi ) &=&-\frac{\mu m\cos \Theta
}{a(1-e^{2})D_{L}}\Bigl[
\left( 5e\sin \phi +4\sin 2\phi +e\sin 3\phi \right) \cos 2\gamma   \notag \\
&&-\left( 5e\cos \phi +4\cos 2\phi +e\cos 3\phi +2e^{2}\right) \sin
2\gamma
\Bigr]\,,  \label{classhx} \\
h_{+}(\phi ) &=&-\frac{\mu m\left( 1+\cos ^{2}\Theta \right) }{
a(1-e^{2})D_{L}}\Biggl[\left( \frac{5e}{2}\cos \phi +2\cos 2\phi
+\frac{e}{2}
\cos 3\phi +e^{2}\right) \cos 2\gamma   \notag \\
&&+\left( \frac{5e}{2}\sin \phi +2\sin 2\phi +\frac{e}{2}\sin 3\phi
\right) \sin 2\gamma +\left( e\cos \phi +e^{2}\right) \frac{\sin
^{2}\Theta }{1+\cos ^{2}\Theta }\Biggr]\,.  \label{classh+}
\end{eqnarray}
\end{widetext}
Here $\phi $ is the true anomaly, which describes the azimuthal
angle from the pericenter along the orbit as shown in Fig.
\ref{geometry}. The value $\gamma $ is the azimuthal angle of the
pericenter relative to the coordinate system $x$ axis in the orbital
plane, $e$ is the orbital eccentricity, $a$ is the semimajor axis,
$D_{L}$ is the luminosity distance, $\Theta $ is the inclination
(the
angle between the orbital plane and the line of sight to the observer), and~$%
m=m_{1}+m_{2}$, $\mu =m_{1}m_{2}/m$ are the total and reduced masses (Fig.%
\ref{geometry}). Using the well-known Fourier-Bessel decomposition, the
polarization states can be expressed as a sum of harmonics of the orbital
frequency \cite{ecc1}
\begin{eqnarray}
\widetilde{h}_{\times }(t) &=&-h\cos \Theta \underset{n}{\sum }\left[
B_{n}^{-}\sin \Phi _{n+}^{t}+B_{n}^{+}\sin \Phi _{n-}^{t}\right] \ ,
\label{hxt} \\
\widetilde{h}_{+}(t) &=&-\tfrac{h}{2}\underset{n}{\sum }\Bigl[\sin
^{2}\Theta A_{n}\cos \Phi _{n}^{t}  \notag \\
&&+\left( 1+\cos ^{2}\Theta \right) \left( B_{n}^{+}\cos \Phi
_{n-}^{t}-B_{n}^{-}\cos \Phi _{n+}^{t}\right) \Bigr]\ .  \label{h+t}
\end{eqnarray}
Here $h=4\mu m(aD_{L})^{-1}$ is the amplitude, and $B_{n}^{\pm
}=\left( S_{n}\pm C_{n}\right) /2$ and $A_{n}$ are linear
combinations of the Bessel functions of the first kind [$J_{n}(ne)$]
and their derivatives,
\begin{eqnarray}
S_{n} &=&-\frac{2\left( 1-e^{2}\right) ^{1/2}}{e}n^{-1}J_{n}^{\prime }(ne)+
\frac{2\left( 1-e^{2}\right) ^{3/2}}{e^{2}}nJ_{n}(ne)\ ,  \notag \\
C_{n} &=&-\frac{2-e^{2}}{e^{2}}J_{n}(ne)+\frac{2\left( 1-e^{2}\right) }{e}
J_{n}^{\prime }(ne)\ ,  \notag \\
A_{n} &=&J_{n}(ne)\ \ ,
\end{eqnarray}
where a prime denotes the derivative, i.e. $J_{n}^{\prime
}(ne)\equiv n\left[ J_{n-1}(ne)+J_{n+1}(ne)\right] /2$. The phase
functions in Eqs.~(\ref{hxt}--\ref{h+t}) are
\begin{eqnarray}
\Phi _{n}^{t} &=&nl\ ,  \label{t1} \\
\Phi _{n\pm }^{t} &=&nl\pm 2\gamma \ ,  \label{t2}
\end{eqnarray}
where $l$ is the mean anomaly which is defined by the Kepler equation
\begin{equation}
l=\xi -e\sin \xi =2\pi \nu (t-t_{0})\ .  \label{Kepler}
\end{equation}
In the Kepler equation $\xi $ is the eccentric anomaly, $\nu
=T^{-1}$ is the Keplerian orbital frequency (here $T=2\pi
m^{-1/2}a^{3/2}$ is the Newtonian radial orbital period), and
$t_{0}$ is the time of pericenter passage (Hereafter, we set
$t_{0}=0$.) Equations~(\ref{t1},\ref{t2}) show that the phase splits
into a triplet due to the pericenter position $\gamma $. If the
pericenter precesses, a triplet of frequencies appears in Fourier
space for each harmonic \cite{ecc1,ecc2}. Note that
Eq.~(\ref{Kepler}) is approximately valid during an orbit as long as
$v/c\ll 1$ and $\nu=\mathrm{constant}$, but this equation requires
modifications on large time scales where the binary inspirals (see
Eqs.~\ref{phase1}--\ref{phase2} below), or at small separations,
where the 1 PN treatment breaks down.

Pericenter precession leads to a time-dependent angle of pericenter,
which may be written as $\gamma (t)=\gamma _{0}+\gamma (t)$ where
$\gamma _{0}$ is the initial angle of the pericenter
(Fig.~\ref{geometry}).
Henceforth, we adopt pericenter precession from the classical
relativistic motion and assume the adiabatic evolution of the
orbital parameters. These effects are averaged over one radial
oscillation period, i.e. $\left\langle \dot{\gamma}\right\rangle
=\Delta \gamma /T$, where $\Delta \gamma =6\pi m[a(1-e^{2})]^{-1}$
is the angle of precession for an eccentric orbit governed by the
geodesic equation of the Schwarzschild geometry (see e.g. Ref.
\cite{Kidder}). In the following we shall drop $\left\langle
{}\right\rangle $ for the average quantities, so we write
\begin{equation}
\dot{\gamma}=\frac{3m^{3/2}}{a^{5/2}(1-e^{2})}=\frac{3m^{2/3}\left( 2\pi \nu
\right) ^{5/3}}{(1-e^{2})}\ .  \label{gammadot}
\end{equation}
The 2.5 PN leading-order adiabatic evolution of the orbital
parameters due to gravitational radiation averaged over one radial
period are \cite{P}
\begin{eqnarray}
\dot{\nu} &=&\frac{48\mathcal{M}^{5/3}(2\pi \nu )^{11/3}}{5\pi
(1-e^{2})^{7/2}}\left( 1+\frac{73}{24}e^{2}+\frac{37}{96}e^{4}\right) \ ,
\label{PP1} \\
\dot{e} &=&-\frac{304\mathcal{M}^{5/3}(2\pi \nu )^{8/3}}{15(1-e^{2})^{5/2}}%
e\left( 1+\frac{121}{304}e^{2}\right) \ ,  \label{PP2}
\end{eqnarray}
where $\mathcal{M}=\mu ^{3/5}m^{2/5}$ is the chirp mass. (We used
Kepler's third law, i.e. $\nu =\left( 2\pi \right)
^{-1}m^{1/2}a^{-3/2}$.)

For an inspiraling system, the phase functions are $\Phi
_{n}^{t}=2\pi n\int_{-\infty }^{t}\nu (t^{\prime })dt^{\prime }$ and
$\Phi _{n\pm }^{t}=\Phi _{n}\pm 2\gamma _{0}\pm 2\int_{-\infty
}^{t}\dot{\gamma} (t^{\prime })dt^{\prime }$. Equations (\ref{t1}),
(\ref{t2}), are generalized as (here the $t$ index is suppressed in
$\Phi _{n}^{t},\Phi _{n\pm }^{t}$)
\begin{eqnarray}
\Phi _{n} &=&2\pi n\int_{-\infty }^{\nu (t)}\frac{\nu}{\dot{\nu}}d\nu \ ,
\label{phase1} \\
\Phi _{n\pm } &=&\Phi _{n}\pm 2\gamma _{0}\pm 2\int_{-\infty }^{\nu (t)}
\frac{\dot{\gamma} }{\dot{\nu}}d\nu \ ,  \label{phase2}
\end{eqnarray}
where $\Phi _{n\pm }$ are phase functions which arise due to
pericenter precession. Note that here one must incorporate the
evolution in the eccentricity by solving Eqs.~(\ref{PP1},\ref{PP2}),
i.e. $\dot\nu\equiv \dot{\nu}(\nu) = \dot\nu[\nu,e(\nu)]$, and
similarly for $\dot{\gamma}$ [see Eq.~(\ref{e:nu(e)}) below].

\section{Fourier transformation of the eccentric inspiral waveform}

The sensitivity of a GW detector is usually given in Fourier space.
Thus, to estimate the detection signal-to-noise ratio and
measurement accuracy, we construct the Fourier transform of the
waveform as
\begin{equation}
h(f)=\underset{-\infty }{\overset{\infty }{\int }}\widetilde{h}(t)e^{2\pi
itf}dt\ ,  \label{Fourier}
\end{equation}
where $f$ is the Fourier frequency. These integrals cannot be evaluated
analytically without further assumptions. However, since the orbital
parameters ($a,e$) evolve very slowly relative to the GW phase, the
stationary phase approximation (SPA) can be utilized \cite{ecc2} (Appendix
B). We account for the adiabatic time evolution during the inspiral in the
Fourier-transformed waveform $h(f)$ using Eqs.~(\ref{phase1}--\ref{phase2})
and the SPA. In this approximation the Fourier transformation of the
waveform becomes a discrete sum over the harmonics of orbital frequency, $%
f_{n}=n\nu $. When the pericenter precession is taken into account, each
harmonic $f_{n}$, is split into a triplet $\mathbf{f\equiv }(f_{n},f_{n\pm
}) $ and therefore the waveform consists of the sum over these triplets of
Fourier frequencies:
\begin{eqnarray}
h_{\times }(\mathbf{f}) &=&-\frac{h_{0}}{2}\underset{n}{\sum }\cos \Theta%
\Bigl[B_{n}^{-}\Lambda _{+}e^{i\left( \Psi _{n+}+\pi /4\right) }  \notag \\
&&+B_{n}^{+}\Lambda _{-}e^{i(\Psi _{n-}+\pi /4)}\Bigr]\ ,  \label{elll1} \\
h_{+}(\mathbf{f}) &=&-\frac{h_{0}}{4}\underset{n}{\sum }\Bigl[\sin
^{2}\Theta A_{n}\Lambda e^{i(\Psi _{n}-\pi /4)}  \notag \\
&&+\left( 1+\cos ^{2}\Theta \right) \Bigl(B_{n}^{+}\Lambda _{-}e^{i(\Psi
_{n-}-\pi /4)}  \notag \\
&&-B_{n}^{-}\Lambda _{+}e^{i(\Psi _{n+}-\pi /4)}\Bigr)\Bigr]\ ,
\label{elll2}
\end{eqnarray}
where $f_{n}=n\nu$, $f_{n\pm }=n\nu \pm \frac{\dot{\gamma}}{\pi }$;
$h_{0}=4\mathcal{M}^{5/3}\left( 2\pi \nu \right) ^{2/3}/D_{L}$ is
the amplitude corresponding to the orbital frequency; and $\Psi
_{n}=2\pi ft_{n}-\Phi _{n}$, $\Psi _{n\pm }=2\pi ft_{n\pm }-\Phi
_{n\pm }$ are phase functions (where $t_{n}$, $t_{n\pm }$ are the
time parameters of the SPA; see Appendix B). We have introduced the
notation $\Lambda _{\pm }=(n\dot{\nu}\pm \ddot{\gamma}/\pi )^{-1/2}$
and $\Lambda =\left( n\dot{\nu}\right) ^{-1/2}$. The phases $\Psi
_{n}$ and $\Psi _{n\pm }$ depend on the corresponding Fourier
frequencies $f_{n}$, $f_{n\pm }$, respectively.

We recall that for circular orbits (i.e. $e\rightarrow 0$) the
waveforms in Eqs. (\ref{elll1},\ref{elll2}) simplify as
\begin{eqnarray}
h_{\times }^{\circ }(f) &=&-2\sqrt{\frac{5}{96}}\frac{\mathcal{M}
^{5/6}f^{-7/6}}{\pi ^{2/3}D_{L}}\cos \Theta e^{i\Psi _{\circ }^{+}}\ , \\
h_{+}^{\circ }(f) &=&-\sqrt{\frac{5}{96}}\frac{\mathcal{M}^{5/6}f^{-7/6}}{
\pi ^{2/3}D_{L}}\left( 1+\cos ^{2}\Theta \right) e^{i\Psi _{\circ }^{-}}\ ,
\end{eqnarray}
where $f=2\nu $ is the (circular) Fourier frequency, and $\Psi
_{\circ }^{\pm }=2\pi ft_{c}-\Phi _{c}\pm \pi /4+\left( 3/4\right)
(8\pi \mathcal{M} f)^{-5/3}$ is the well-known phase function.

\subsection{LISA detector response}

With its three arms, LISA represents a pair of two orthogonal arm
detectors, $I$ and $II$, producing two linearly independent signals.
The frequency domain waveforms are
\begin{equation}
h^{I,II}(\mathbf{f})=\frac{\sqrt{3}}{2}\left[ F_{\times }^{I,II}h_{\times
}\left( \mathbf{f}\right) +F_{+}^{I,II}h_{+}\left( \mathbf{f}\right) \right]
\ ,  \label{hF}
\end{equation}
with the antenna beam pattern functions
\begin{eqnarray}
F_{\times }^{I} &=&\tfrac{1+\mu _{S}^{2}}{2}\cos 2\phi _{S}\sin 2\psi
_{S}+\mu _{S}\sin 2\phi _{S}\cos 2\psi _{S}\ , \\
F_{+}^{I} &=&\tfrac{1+\mu _{S}^{2}}{2}\cos 2\phi _{S}\cos 2\psi _{S}-\mu
_{S}\sin 2\phi _{S}\sin 2\psi _{S}\ ,
\end{eqnarray}
where $\mu _{S,L}=\cos \theta _{S,L}$ with ($\theta _{S},\phi _{S}$)
being spherical angles of the source in the detector-based
coordinate system. The angle $\psi _{S}$ is the polarization angle
that can be expressed by the position of the detector and the
orbital plane \cite{Cutler98}. The other antenna beam pattern
functions are $F_{+,\times }^{II}=F_{+,\times }^{I}(\phi _{S}-\pi
/4)$. The quantities $\theta _{S}$, $\phi _{S}$, and $\psi _{S}$ are
time dependent, because the LISA constellation moves around the Sun,
and these explicit time evolutions are \cite{Cutler98}
\begin{eqnarray}
\mu _{S} &=&\tfrac{\bar{\mu}_{S}}{2}-\tfrac{\sqrt{3}\bar{\lambda}_{S}}{2}%
\cos \bar{\phi}_{S}^{t}\ ,  \label{pol1} \\
\phi _{S} &=&\alpha _{1}(t)+\tfrac{\pi }{12}+\arctan \tfrac{\sqrt{3}\bar{\mu}%
_{S}+\bar{\lambda}_{S}\cos \bar{\phi}_{S}^{t}}{2\bar{\lambda}_{S}\sin \bar{%
\phi}_{S}^{t}}\ ,  \label{pol2} \\
\psi _{S} &=&\arctan \tfrac{\bar{\mu}_{L}-\sqrt{3}\bar{\lambda}_{L}\cos \bar{%
\phi}_{L}^{t}-\cos \Theta \left( \bar{\mu}_{S}-\sqrt{3}\bar{\lambda}_{S}\cos
\bar{\phi}_{S}^{t}\right) }{2K}\ ,  \label{pol3}
\end{eqnarray}%
where $\bar{\lambda}_{S,L}=\sin \bar{\theta}_{S,L}$,
$\bar{\mu}_{S,L}=\cos \bar{\theta}_{S,L}$ and
$\bar{\phi}_{S,L}^{t}=\bar{\phi}(t)-\bar{\phi}_{S,L}$, with
$\bar{\theta}_{S}$, $\bar{\phi}_{S}$ being the spherical angles of
the source's position. The angles $\bar{\theta}_{L}$,
$\bar{\phi}_{L}$ correspond to the direction of orbital angular
momentum in the barycenter frame \cite{Cutler98}. In Eqs.
(\ref{pol1}-\ref{pol3}), $\Theta =\arccos \left[
\bar{\mu}_{L}\bar{\mu}_{S}+\bar{\lambda}_{L}\bar{\lambda}_{S}\cos
(\bar{\phi}_{L}-\bar{\phi}_{S})\right] $ is the inclination [in Eqs.
(\ref{hxt},\ref{h+t})] and the explicit time dependences are $\alpha
_{1}(t)=2\pi t/T-\pi /12+\alpha _{0}$,
$\bar{\phi}(t)=\bar{\phi}_{0}+2\pi t/T $, and
\begin{eqnarray}
K &=&\tfrac{\bar{\lambda}_{L}\bar{\lambda}_{S}}{2}\sin (\bar{\phi}_{L}-\bar{\phi}_{S})  \notag \\
&&-\tfrac{\sqrt{3}}{2}\cos \bar{\phi}(t)\left( \bar{\mu}_{L}\lambda _{S}\sin
\bar{\phi}_{S}-\bar{\mu}_{S}\bar{\lambda}_{L}\sin \bar{\phi}_{L}\right)
\notag \\
&&-\tfrac{\sqrt{3}}{2}\sin \bar{\phi}(t)\left(
\bar{\mu}_{S}\bar{\lambda}_{L}\cos
\bar{\phi}_{L}-\bar{\mu}_{L}\bar{\lambda}_{S}\cos
\bar{\phi}_{S}\right)\,.
\end{eqnarray}
We note that $\bar{\theta}_{L}$, $\bar{\phi}_{L}$ are generally not
constants for spinning binaries due to spin-orbit effects, but we neglect
these effects here.

We carry out the analysis for the single-detector case ($I$ only) and the
full two-detector configuration ($I+II$).

In practice, the measured signal in Eq. (\ref{hF}) is truncated at
minimum and maximum frequencies corresponding to the start of the
observation and the last stable orbit for each harmonic,
respectively (see Sec. V below).

\section{Parameter estimation}

In this section we review the basics of Bayesian parameter estimation. The
measured signal $\widetilde{s}(t)$ is made up of the GW $\widetilde{h}(t)$
and the noise $\widetilde{n}(t)$
\begin{equation}
\widetilde{s}(t)=\widetilde{h}(t)+\widetilde{n}(t)\ .
\end{equation}
We assume that the noise is stationary, Gaussian, and statistically
independent at different frequencies. Then each Fourier component
has a Gaussian probability distribution and the different Fourier
components of the noise are "uncorrelated." i.e.
\begin{eqnarray}
p(n &=&n_{0})\propto e^{-(n_{0}\mid n_{0})^{2}}\ ,  \label{noise1s} \\
\left\langle n(f)n^{\ast }(f^{\prime })\right\rangle &=&\tfrac{1}{2}\delta
(f-f^{\prime })S(f)\ .  \label{noise2s}
\end{eqnarray}
In Eqs. (\ref{noise1s},\ref{noise2s}) $p(n)$ is the probability for
the noise, the inner product is defined by
\begin{equation}
(g\mid k)=4\Re \int_{0}^{\infty }\frac{g(f)k^{\ast }(f)}{S(f)}df\ ,
\end{equation}
$k^{\ast }$ is denotes complex conjugation and $S(f)$ is the
one-sided spectral noise density. The definition of the
signal-to-noise ratio (SNR) of $h$ is
\begin{equation}
\rho ^{2}=(h\mid h)=4\Re \underset{0}{\overset{\infty }{\int }}\frac{%
h(f)h^{\ast }(f)}{S(f)}df\ \ .  \label{SNR2}
\end{equation}
The waveform $h(f)$ depends on the parameters $\lambda ^{a}$ which
characterize the source. For a large SNR, the errors $\Delta \lambda
^{a}$ have the Gaussian probability distribution
\begin{equation}
p(\Delta \lambda ^{c})=p_{0}e^{-\Gamma _{ab}\Delta \lambda ^{a}\Delta
\lambda ^{b}/2}\ .
\end{equation}
where $p_{0}$ is the normalization factor and $\Gamma _{ab}$ is the Fisher
information matrix defined by
\begin{equation}
\mathbf{\Gamma }_{ab}=(\partial _{a}h\mid \partial _{b}h)=4\Re
\underset{0}{\overset{\infty }{\int }}\frac{\partial
_{a}h(f)\partial _{b}h^{\ast }(f)}{S(f)}df\ \ ,  \label{fisher}
\end{equation}
with $\partial _{a}=\partial /\partial \lambda ^{a}$. The inverse of
the Fisher matrix is approximately the $\Sigma _{ab}$
variance-covariance matrix for $\rho \gg 1$, which gives the
accuracy of each parameter and is defined by $\Sigma _{ab}=\left(
\Gamma _{ab}\right) ^{-1}=\left\langle \Delta \lambda ^{a}\Delta
\lambda ^{b}\right\rangle $. The root-mean-square errors of the
parameters $\lambda ^{a}$ are $\Delta \lambda
^{a}=\sqrt{\mathbf{\Sigma }_{aa}}$.

For example, the error of the sky position solid angle is
\begin{equation}
\Delta \Omega _{S}=2\pi \sqrt{\left( \Delta \overline{\mu
}_{S}\Delta \overline{\phi }_{S}\right) ^{2}-\left\langle \Delta
\overline{\mu }_{S}\Delta \overline{\phi }_{S}\right\rangle ^{2}}\ .
\end{equation}

The source localization sky area is an ellipse with semiminor and major axes
$(a_{S},b_{S})$ given by Eq. (4.12) in Ref.~\cite{LH}. The SNR and Fisher
matrix for the LISA configuration are
\begin{eqnarray}
\rho ^{2} &=&\rho _{I}^{2}+\rho _{II}^{2}\ ,  \notag \\
\mathbf{\Gamma }_{ab} &=&\mathbf{\Gamma }_{ab}^{I}+\mathbf{\Gamma }
_{ab}^{II}\ .
\end{eqnarray}
where the $I$, $II$ subscripts distinguish the $h^{I}$, $h^{II}$ waveforms
in Eq. (\ref{hF}).

\section{Measuring eccentric inspiraling SMBH binaries}

We focus on comparable-mass SMBH binaries in the range
$(10^{4}-10^{7})M_{ \odot }$, which corresponds to the measured
frequency range $10^{-4}$ to $10^{-1}$Hz. For initial configurations
1 yr before merger, we assume that the binary has orbital
eccentricity $e_{0}$ and pericenter position $\gamma _{0}$. The
ten-dimensional parameter space is
\begin{equation*}
\lambda ^{a}=\{\ln D_{L},\ln \mathcal{M},t_{c},\Phi _{c},\bar{\phi}_{S},\bar{
\mu}_{S},\bar{\phi}_{L},\bar{\mu}_{L},e_{0},\gamma _{0}\}
\end{equation*}
In the circular case $e_{0}$ and $\gamma _{0}$ do not appear. Note
that only one mass parameter, the chirp mass $\mathcal{M}$, enters
the leading-order waveform. Our assumptions are as follows:

\begin{itemize}
\item[--] To examine the effects of eccentricity and pericenter precession,
we neglect higher-order post-Newtonian (beyond 1 PN) orders and
spins; we only use the \textit{heuristic} pericenter precession in
phase described above.

\item[--] In all cases, we take $t_{c}=\Phi _{c}=\gamma _{0}=0$.
(We use the $\alpha _{0},\bar{\phi}_{0}=0$ choice, as in Ref. \cite{Cutler98}.)

\item[--] We assume that the observation time is 1 yr before the
merger-more precisely, before the \textit{Newtonian} \textit{last
stable orbit} (\textit{LSO}), which is defined by \cite{BC}
\begin{equation}
\nu _{LSO}^{N}=\frac{1}{2\pi m}\left( \frac{1-e_{LSO}^{2}}{6+2e_{LSO}}
\right) ^{3/2}\ ,  \label{cond2}
\end{equation}
where $e_{LSO}$ is the final eccentricity at the last stable orbit
($\nu [e_{LSO})=\nu _{LSO}$].

\item[--] For the $n^{\mathrm{th}}$ orbital harmonic, the limits of
integration are taken to be $\nu _{\max }=\nu _{LSO}$ and $\nu
_{\min }=\max \{\nu _{0},f_{c}/n\}$, where $\nu _{0}$ is the
frequency 1 yr before the \textit{LSO} and
$f_{c}=0.03\,\mathrm{mHz}$ is the cutoff frequency of the LISA
detector.

\item[--] We assume that the luminosity distance to the source is
$D_{L}=6.4\,\mathrm{Gpc}$, corresponding to a cosmological redshift
$z=1$, and we use the comoving masses as free parameters,
$m_{i}^{z}=(1+z)m_{i}$ \cite{Vecchio}. We do not take into account
the Doppler phase due to the varying light travel during the LISA
constellation's: orbit around the Sun.

\item[--] We parametrize the evolution of the orbital frequency with the
instantaneous eccentricity following Ref. \cite{Bence} (Appendix A):

\begin{equation}
\nu (e)=\nu _{0}\frac{\sigma (e)}{\sigma (e_{0})}  \label{e:nu(e)}
\end{equation}
where $\nu _{0}$ and $e_{0}$ are the initial orbital frequency and
eccentricity, and $\sigma (e)$ follows from Ref.~\cite{P}.

\item[--] We truncate the harmonics at $n_{\max }$, where $99\%$ of the
signal power corresponds to \cite{Bence}
\begin{equation}
n_{\max }=\left\lfloor 5\frac{\left( 1+e_{0}\right) ^{1/2}}{\left(
1-e_{0}\right) ^{3/2}}\right\rfloor \ .
\end{equation}
where the bracket $\left\lfloor {}\right\rfloor $ denotes the floor
function (integer part of nonnegative argument). Here $n_{\max
}=\{9,24\}$ for $e_{0}=\{0.3,0.6\}$, respectively.

\item[--] We analyze $10^{4}$ SMBH binaries where the angular variables
were chosen randomly, i.e. for $\bar{\phi}_{S}$, $\bar{\phi}_{L}$ in the
range $(0,2\pi )$ and for $\bar{\theta}_{S}$, $\bar{\theta}_{L}$ in the
range $(-\pi /2,\pi /2)$.
\end{itemize}

The computation of SNR and the Fisher matrix with the above general
definition [Eq. \ref{Fourier}] is numerically expensive for a large
set of binaries. We resort to the SPA waveform. The SNR and the
Fisher information matrix consist of three terms for each orbital
harmonic which correspond to $(f_{n}$, $f_{n\pm })$, respectively:
\begin{eqnarray}
\hat{\rho}^{2} &=&\underset{n}{\sum }\left( \hat{\rho}_{n}^{2}+\hat{\rho}
_{n+}^{2}+\hat{\rho}_{n-}^{2}\right) \\
\mathbf{\hat{\Gamma}}_{ab} &=&\underset{n}{\sum }\left( \mathbf{\hat{\Gamma}}
_{ab}^{n}+\mathbf{\hat{\Gamma}}_{ab}^{n+}+\mathbf{\hat{\Gamma}}
_{ab}^{n-}\right)
\end{eqnarray}
where we have introduced the notations $\hat{\rho}
_{n,n+,n-}^{2}=(h_{n,n+,n-}\mid h_{n,n+,n-})$,
$\mathbf{\hat{\Gamma}}_{ab}^{n,n+,n-}=(\partial _{a}h_{n,n+,n-}\mid
\partial _{b}h_{n,n+,n-})$, and $h_{n,n+,n-}=h(f_{n,n+,n-})$. Here we
neglect the cross terms between different harmonics $n$, $n+$, and
$n-$, in $\hat{\rho}$ and $\mathbf{\hat{\Gamma}}_{ab}$. We use the
LISA sensitivity curve generator \cite{Sh}. In the SPA, we can
change the integration variables from $f_{n}$ ,$f_{n\pm }\,$ to $e$:

\begin{eqnarray}
\hat{\rho}_{n}^{2} &=&4\Re \underset{e_{\min }}{\overset{e_{\max }}{\int }}
\frac{h_{n}(e)h_{n}^{\ast }(e)}{S\left[ n\nu (e)\right] }\frac{nd\nu }{de}
de\ , \\
\mathbf{\hat{\Gamma}}_{ab}^{n} &=&4\Re \underset{e_{\min }}{\overset{e_{\max
}}{\int }}\frac{\partial _{a}h_{n}(e)\partial _{b}h_{n}^{\ast }(e)}{S\left[
n\nu (e)\right] }\frac{nd\nu }{de}de\ ,
\end{eqnarray}
where $d\nu /de$ and $\nu (e)$ are given by Eqs.
(\ref{dnu},\ref{sol2}) and $e_{\max }=e_{LSO}$, $e_{\min }=\min
\{e_{c}(n),e_{0}\}$. (Here $e_{c}(n)$ corresponds to $f_{c}/n$,
where $f_{c}=0.03mHz$ is the cutoff frequency for the LISA
detector.)

\begin{table*}[tb]
\caption{The initial and final frequencies ($\protect\nu _{0}$ and $\protect%
\nu _{1}=\protect\nu _{LSO}$) for various initial eccentricities
($e_{0}$) and comoving masses ($m_1$--$m_2$ with redshift $z=1$) for
a 1 yr inspiral before \textit{LSO}. We use the shorthand notation
$e_{1}=e_{LSO}$ for the final eccentricity. We have completed with a
dimensionless semimajor axis $\bar{r}=a/m$ at the initial
($\bar{r_{0}}$) and final points ($\bar{r_{1}}$).}
\label{table1}%
\begin{equation*}
\begin{tabular}{cccc}
\hline\hline
$SMBH$ $[M_{\odot }]$ & $e_{0}=0$ & $e_{0}=0.3$ & $e_{0}=0.6$ \\ \hline\hline
$10^{7}-10^{7}$ &
\begin{tabular}{c}
$\nu _{0}=3.47\mu \mathrm{Hz},$ $\bar{r}_{0}=37.84$ \\
$\nu _{1}=54.96\mu \mathrm{Hz},$ $\bar{r}_{1}=6.00$%
\end{tabular}
&
\begin{tabular}{c}
$\nu _{0}=3.05\mu \mathrm{Hz},$ $\bar{r}_{0}=41.21$ \\
$\nu _{1}=54.47\mu \mathrm{Hz},$ $\bar{r}_{1}=6.04$ \\
$e_{1}=0.017$%
\end{tabular}
&
\begin{tabular}{c}
$\nu _{0}=1.92\mu \mathrm{Hz},$ $\bar{r}_{0}=56.17$ \\
$\nu _{1}=53.78\mu \mathrm{Hz},$ $\bar{r}_{1}=6.09$ \\
$e_{1}=0.039$%
\end{tabular}
\\ \hline
$10^{6}-10^{6}$ &
\begin{tabular}{c}
$\nu _{0}=14.64\mu \mathrm{Hz},$ $\bar{r}_{0}=67.23$ \\
$\nu _{1}=549.59\mu \mathrm{Hz},$ $\bar{r}_{1}=6.00$%
\end{tabular}
&
\begin{tabular}{c}
$\nu _{0}=12.88\mu \mathrm{Hz},$ $\bar{r}_{0}=73.28$ \\
$\nu _{1}=547.75\mu \mathrm{Hz},$ $\bar{r}_{1}=6.01$ \\
$e_{1}=0.007$%
\end{tabular}
&
\begin{tabular}{c}
$\nu _{0}=8.09\mu \mathrm{Hz},$ $\bar{r}_{0}=99.87$ \\
$\nu _{1}=545.22\mu \mathrm{Hz},$ $\bar{r}_{1}=6.03$ \\
$e_{1}=0.015$%
\end{tabular}
\\ \hline
$10^{5}-10^{5}$ &
\begin{tabular}{c}
$\nu _{0}=61.73\mu \mathrm{Hz},$ $\bar{r}_{0}=119.64$ \\
$\nu _{1}=5495.90\mu \mathrm{Hz},$ $\bar{r}_{1}=6.00$%
\end{tabular}
&
\begin{tabular}{c}
$\nu _{0}=54.31\mu \mathrm{Hz},$ $\bar{r}_{0}=130.30$ \\
$\nu _{1}=5488.93\mu \mathrm{Hz},$ $\bar{r}_{1}=6.01$ \\
$e_{1}=0.003$%
\end{tabular}
&
\begin{tabular}{c}
$\nu _{0}=34.13\mu \mathrm{Hz},$ $\bar{r}_{0}=177.59$ \\
$\nu _{1}=5479.18\mu \mathrm{Hz},$ $\bar{r}_{1}=6.01$ \\
$e_{1}=0.006$%
\end{tabular}
\\ \hline
$10^{4}-10^{4}$ &
\begin{tabular}{c}
$\nu _{0}=260.30\mu \mathrm{Hz},$ $\bar{r}_{0}=212.75$ \\
$\nu _{1}=54959\mu \mathrm{Hz},$ $\bar{r}_{1}=6.00$%
\end{tabular}
&
\begin{tabular}{c}
$\nu _{0}=229.02\mu \mathrm{Hz},$ $\bar{r}_{0}=231.72$ \\
$\nu _{1}=54934\mu \mathrm{Hz},$ $\bar{r}_{1}=6.00$ \\
$e_{1}=0.001$%
\end{tabular}
&
\begin{tabular}{c}
$\nu _{0}=143.94\mu \mathrm{Hz},$ $\bar{r}_{0}=315.80$ \\
$\nu _{1}=54896\mu \mathrm{Hz},$ $\bar{r}_{1}=6.01$ \\
$e_{1}=0.002$%
\end{tabular}
\\ \hline\hline
\end{tabular}%
\end{equation*}%
\end{table*}

\begin{figure}[tbh]
\begin{center}
\includegraphics[width=0.45\textwidth]{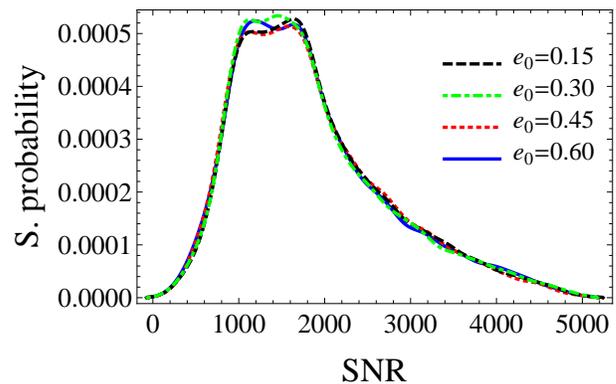}
\end{center}
\caption{(color online) Smooth probability density function of SNR for
various initial eccentricities $e_{0}=0.15,0.3,0.45,0.6$ and masses $\left(
10^{6}-10^{6}\right) M_{\odot }$. The eccentricity dependence of SNR is
almost negligible.}
\label{SNRe}
\end{figure}

\begin{figure}[tbh]
\begin{center}
\includegraphics[width=0.45\textwidth]{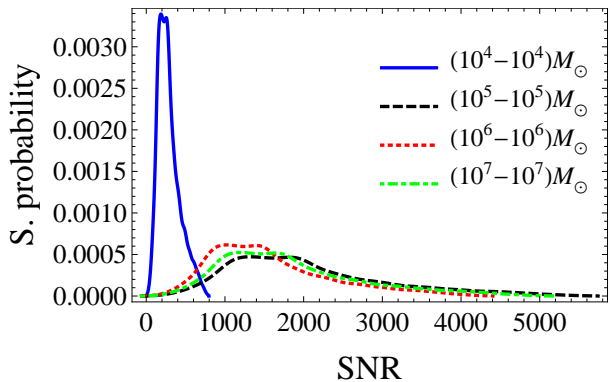}
\end{center}
\caption{(color online). Smooth probability density function of SNR
for various equal-mass binaries (for initial eccentricity
$e_{0}=0.3$). The SNR is $\mathcal{O}(10^{2})$ for low-mass binaries
$\left( 10^{4}-10^{4}\right) M_{\odot }$. In the other cases, the
SNR is $\mathcal{O} (10^{3})$.} \label{SNRm}
\end{figure}

\section{Results and Discussion}

\begin{figure}[t]
\begin{center}
\includegraphics[width=0.52\textwidth]{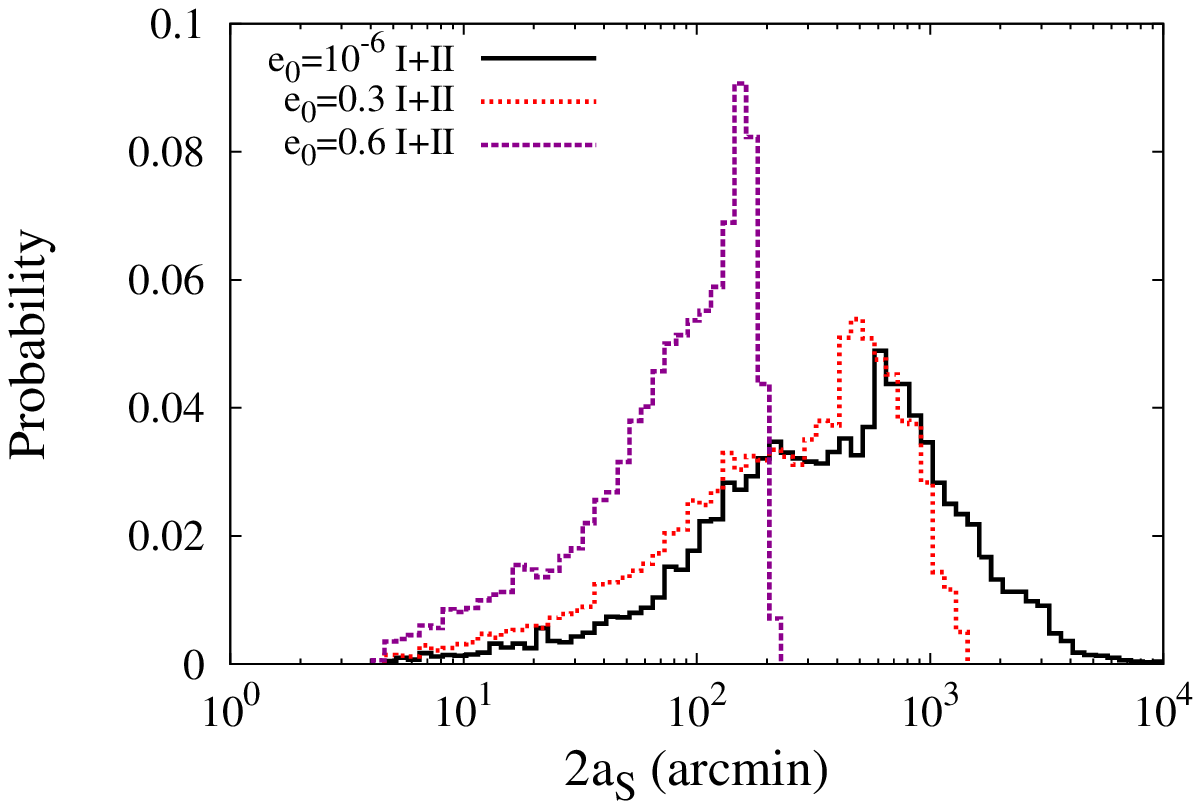} %
\includegraphics[width=0.52\textwidth]{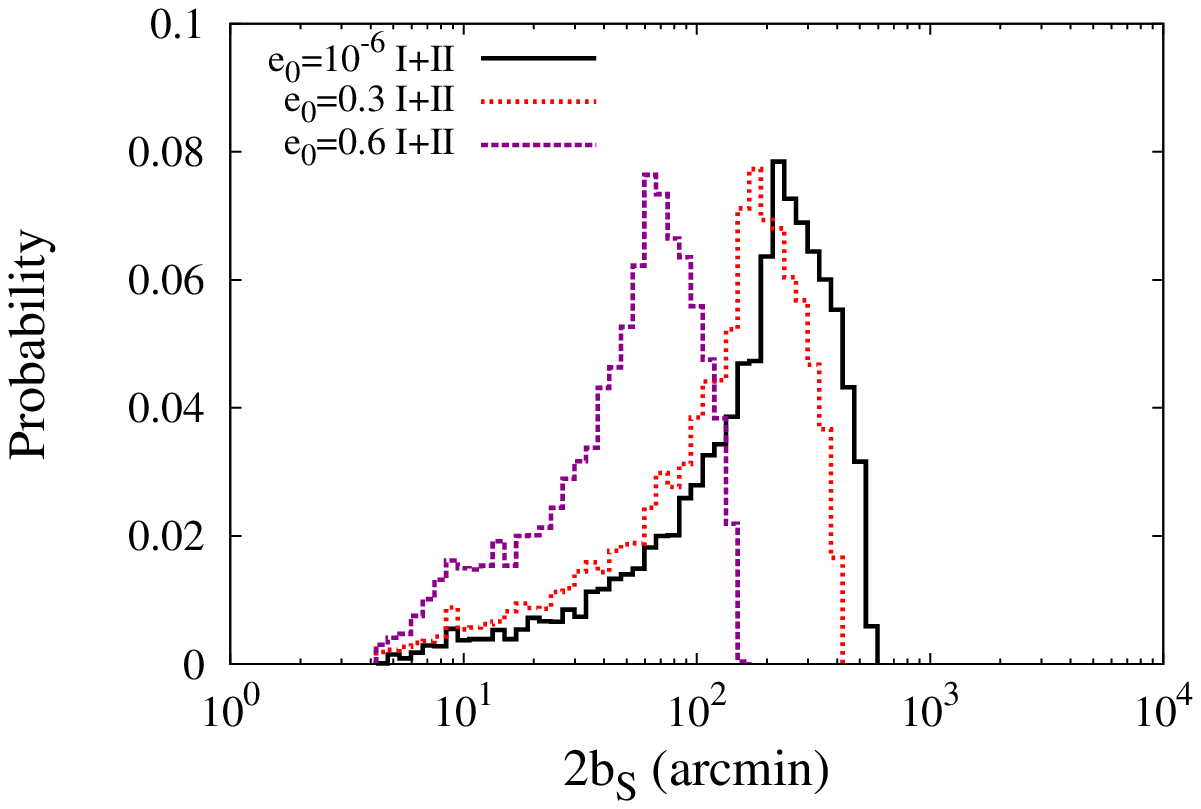}
\end{center}
\caption{(color online). Distribution of the major (top) and minor
(bottom) axes $\left( a_{S},b_{S}\right) $ of the sky position error
ellipse ($\Delta \Omega _{S}=\protect\pi a_{S}b_{S}$) for various
eccentric binaries with equal mass. (Here the pericenter precession
is neglected.) The two panels correspond to 1 yr observation of
$\left( 10^{7}-10^{7}\right) M_{\odot }$ black hole binaries at
$z=1$ ($D_{L}=6.4\mathrm{Gpc}$) with LISA (2 detector). The angular
resolution is improved for high-mass binaries.} \label{Fig3b}
\end{figure}

\begin{figure}[tbh]
\begin{center}
\includegraphics[width=0.52\textwidth]{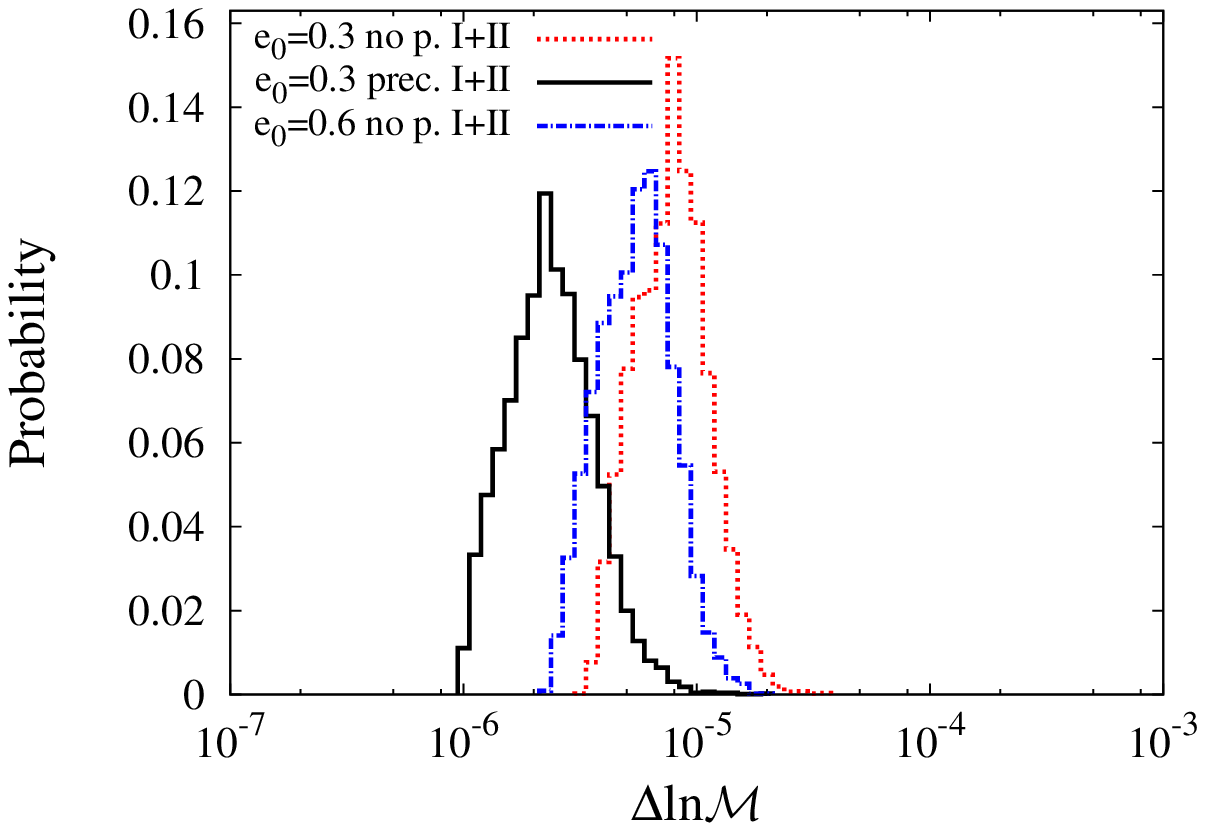}
\includegraphics[width=0.52\textwidth]{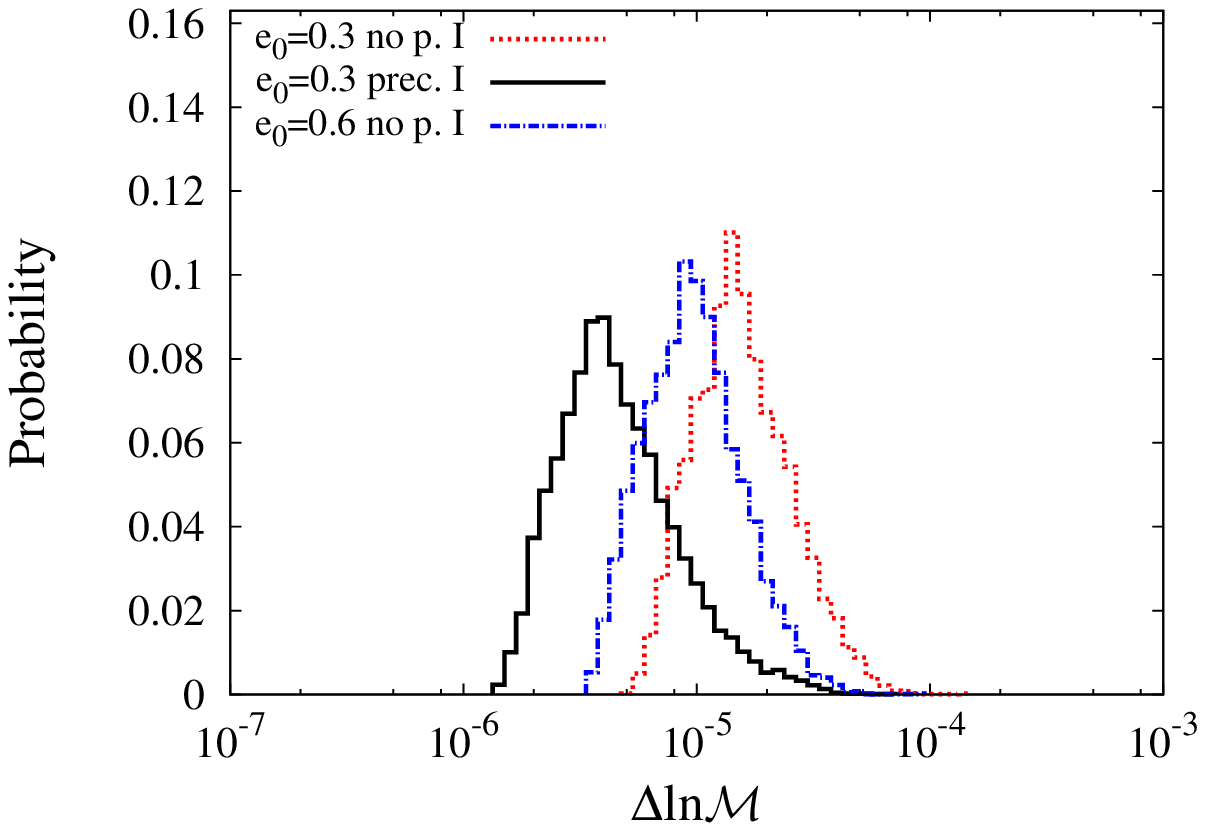}
\end{center}
\caption{(color online). Estimated distribution of the chirp mass
errors in the precessing and nonprecessing cases for the total
($I+II$, top) and single ($I$, bottom) detectors. The results are
shown for medium ($e_{0}=0.3$) and high ($e_{0}=0.6$) initial
eccentricities and higher-mass SMBH binaries $\left(
10^{6}-10^{6}\right) M_{\odot }$. For precessing sources the
$e_{0}=0.6$ case is omitted in both figures due to the high degree
of overlap with the $e_{0}=0.3$ case.} \label{chirp66}
\end{figure}

\begin{figure}[tbh]
\begin{center}
\includegraphics[width=0.52\textwidth]{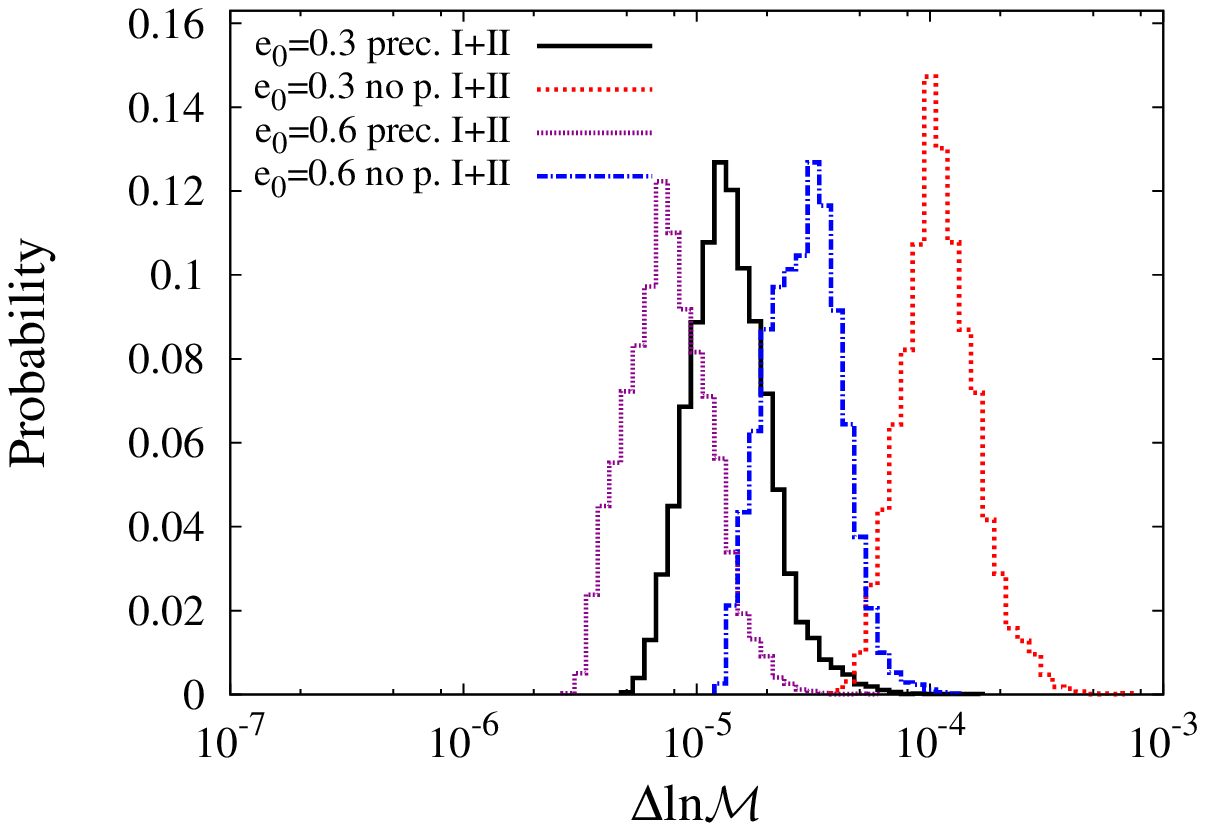}
\includegraphics[width=0.52\textwidth]{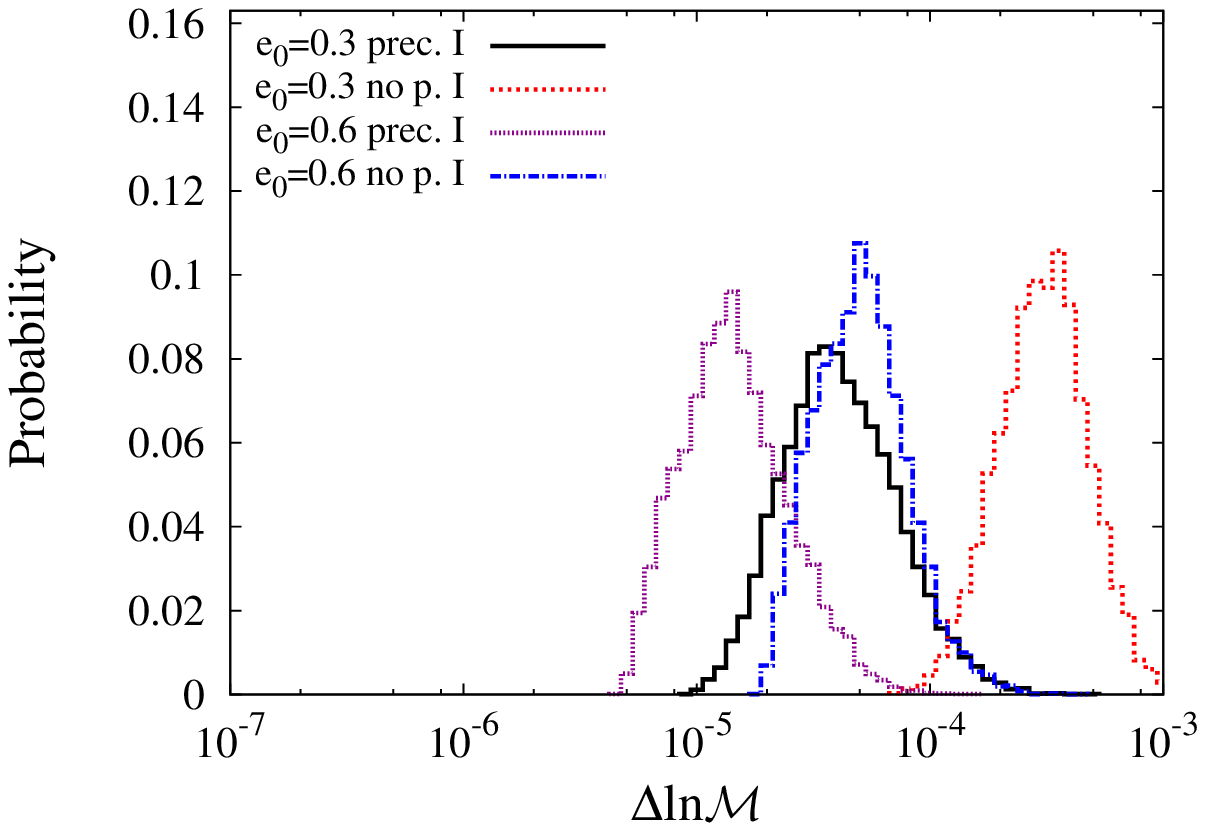}
\end{center}
\caption{(color online). Same as Fig.~\protect\ref{chirp66} but for
masses $\left( 10^{7}-10^{7}\right) M_{\odot }$.} \label{chirp77}
\end{figure}

\begin{figure}[tbh]
\begin{center}
\includegraphics[width=0.52\textwidth]{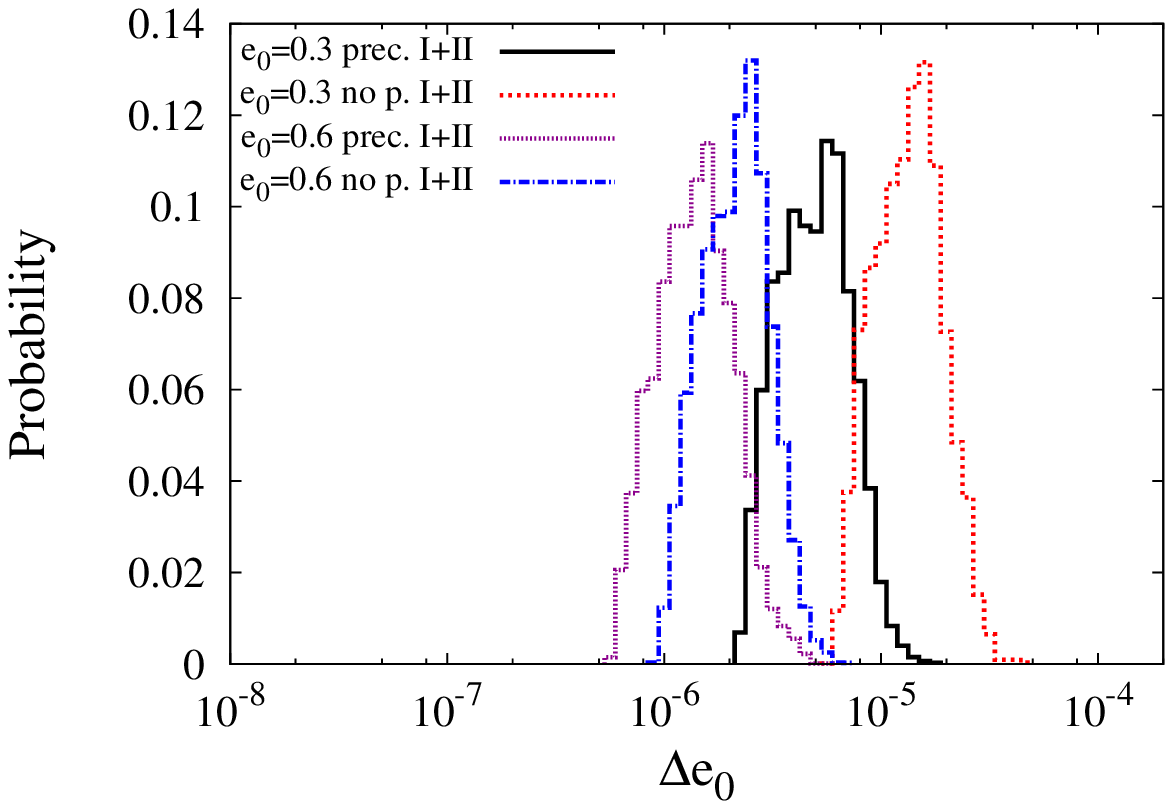}
\includegraphics[width=0.52\textwidth]{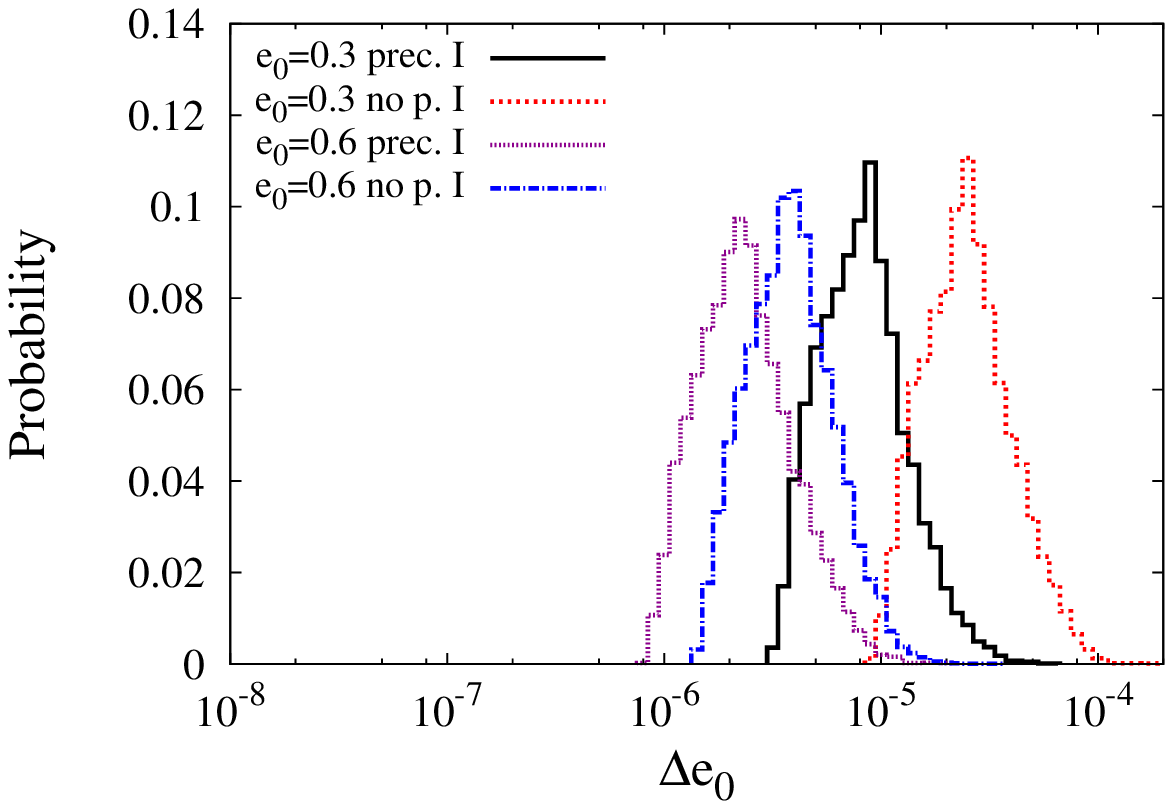}
\end{center}
\caption{(color online). Estimated distribution of the initial
eccentricity errors in the precessing and nonprecessing cases for
the total ($I+II$, top) and single ($I$, bottom) detectors. The
results are shown for medium ($e_{0}=0.3$) and high ($e_{0}=0.6$)
initial eccentricities and higher-mass SMBH binaries $\left(
10^{6}-10^{6}\right) M_{\odot }$.} \label{e066}
\end{figure}

\begin{figure}[tbh]
\begin{center}
\includegraphics[width=0.52\textwidth]{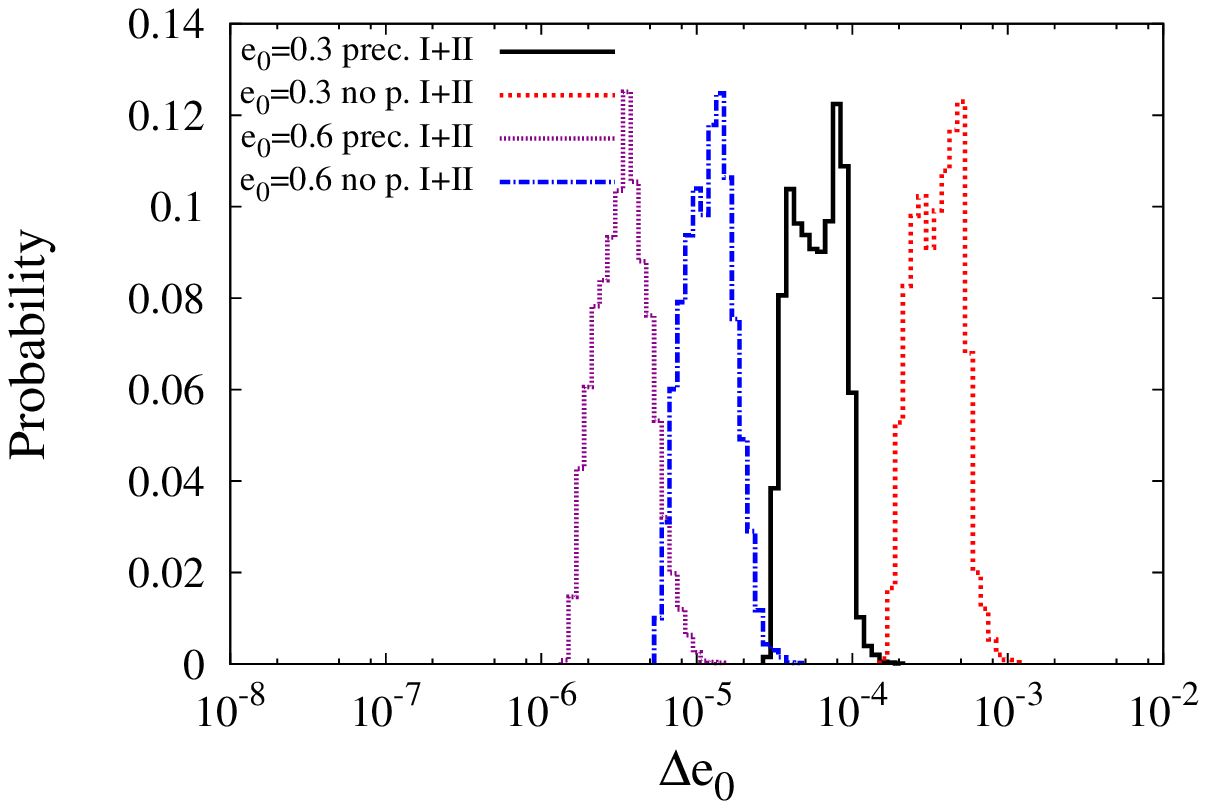}
\includegraphics[width=0.52\textwidth]{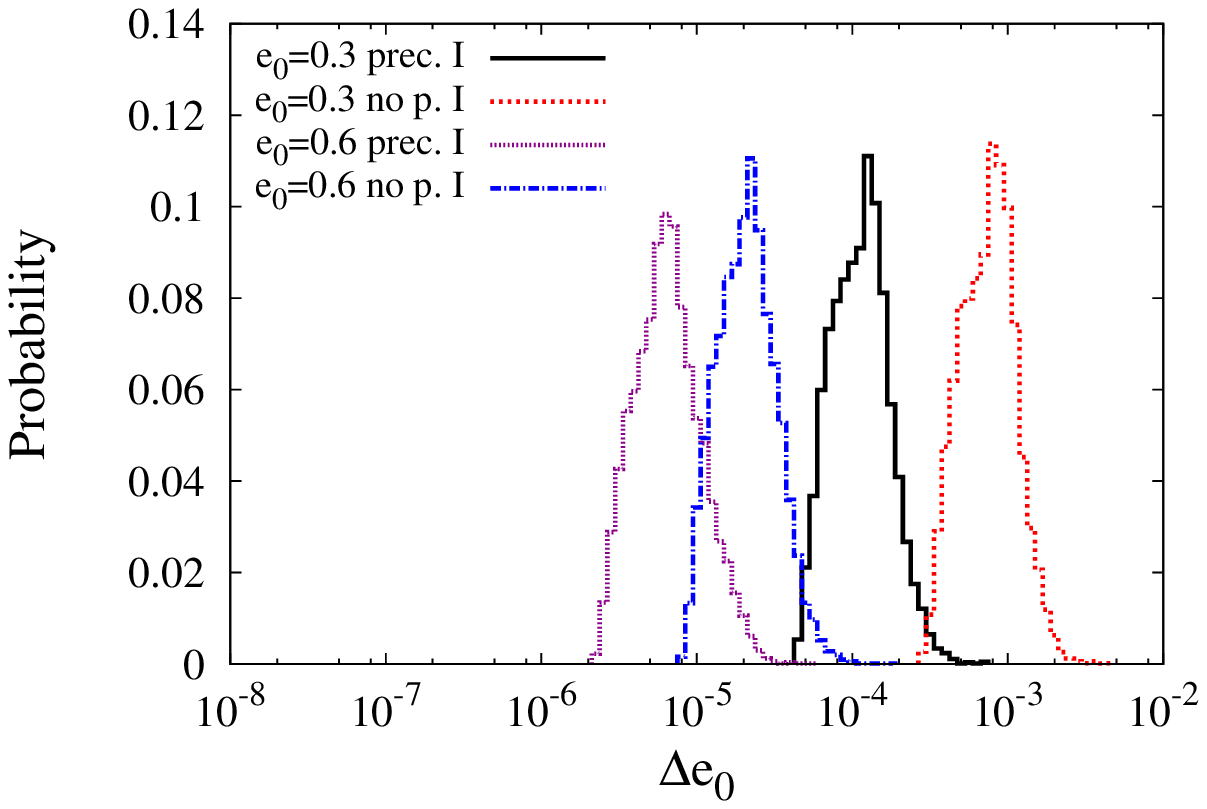}
\end{center}
\caption{(color online). Same as Fig.~\protect\ref{e066} but for
masses $\left( 10^{7}-10^{7}\right) M_{\odot }$.} \label{e077}
\end{figure}

\begin{figure}[tbh]
\begin{center}
\includegraphics[width=0.52\textwidth]{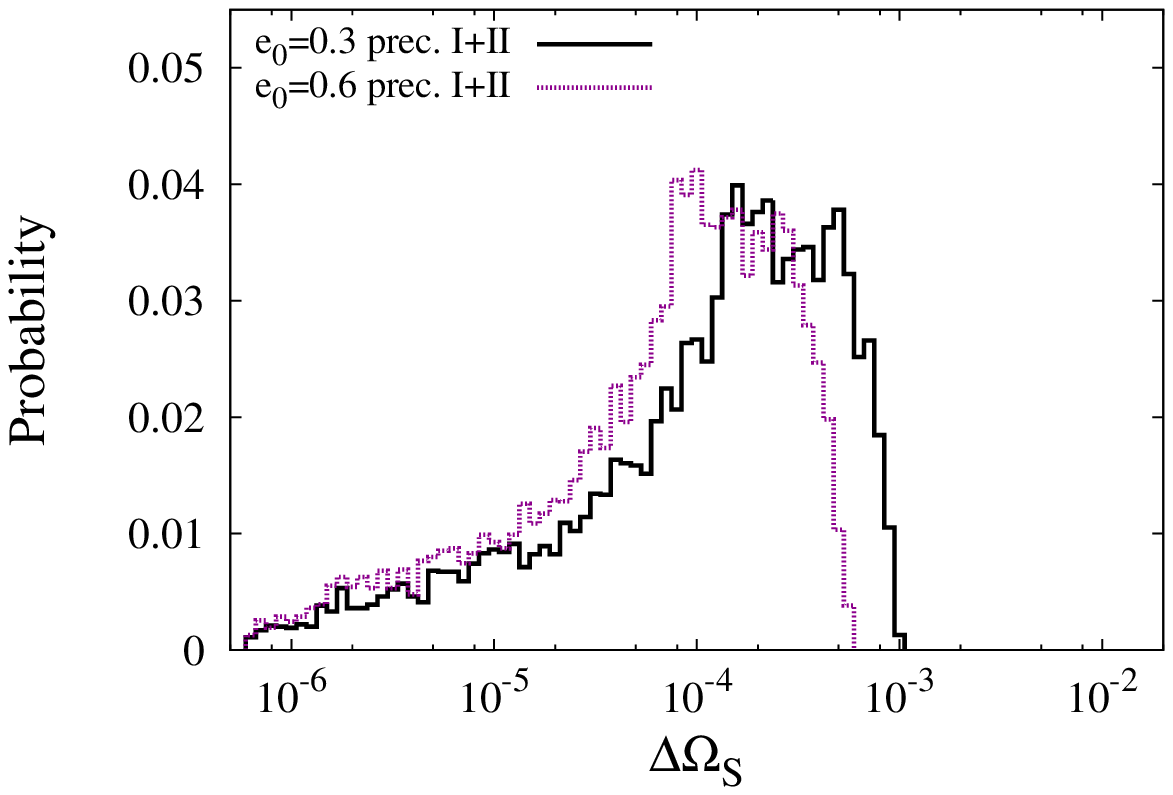}
\includegraphics[width=0.52\textwidth]{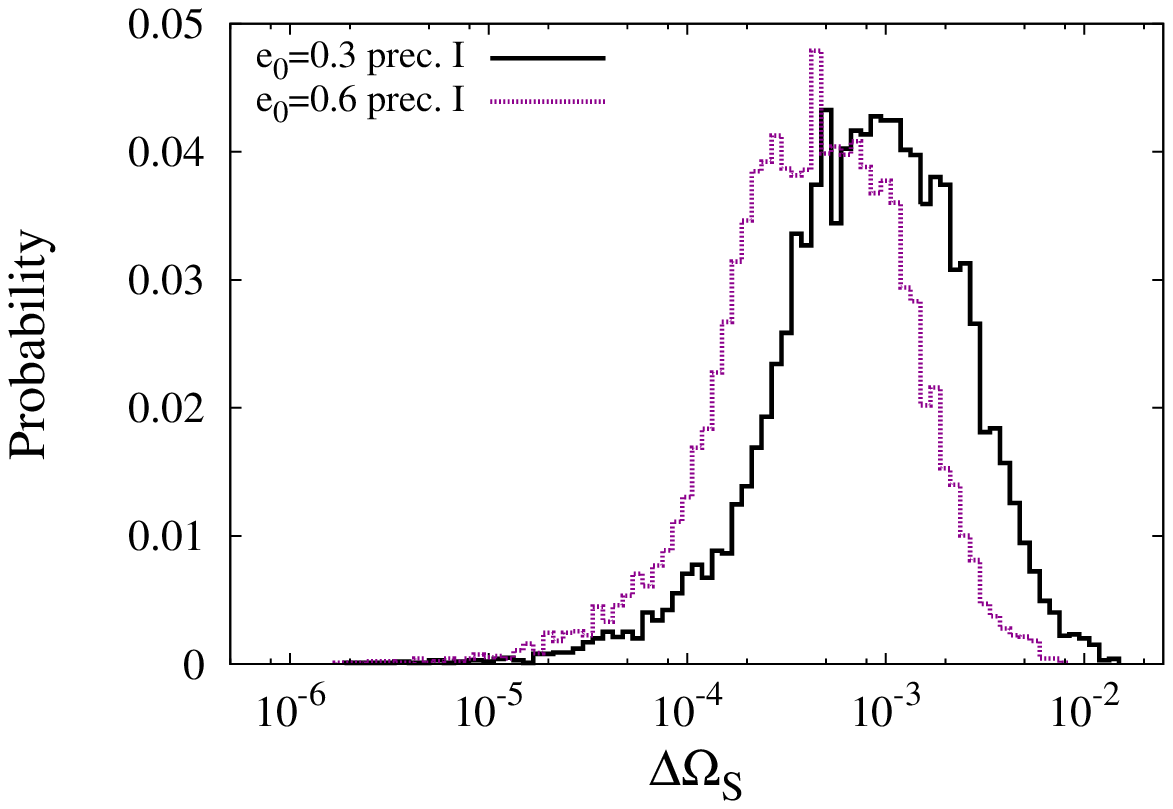}
\end{center}
\caption{(color online). Estimated distribution of the angular
resolution $\Delta\Omega_{S}$ in the precessing case for the total
($I+II$, top) and single ($I$, bottom) detectors. The results are
shown for medium ($e_{0}=0.3$) and high ($e_{0}=0.6$) initial
eccentricities and higher-mass SMBH binaries $\left(
10^{6}-10^{6}\right) M_{\odot }$. The curves for the nonprecessing
$e_{0}=0.3$ and $e_{0}=0.6$ cases are omitted in both figures since
they are close and very similar to the curves for the precessing
ones.} \label{angular66}
\end{figure}

\begin{figure}[tbh]
\begin{center}
\includegraphics[width=0.52\textwidth]{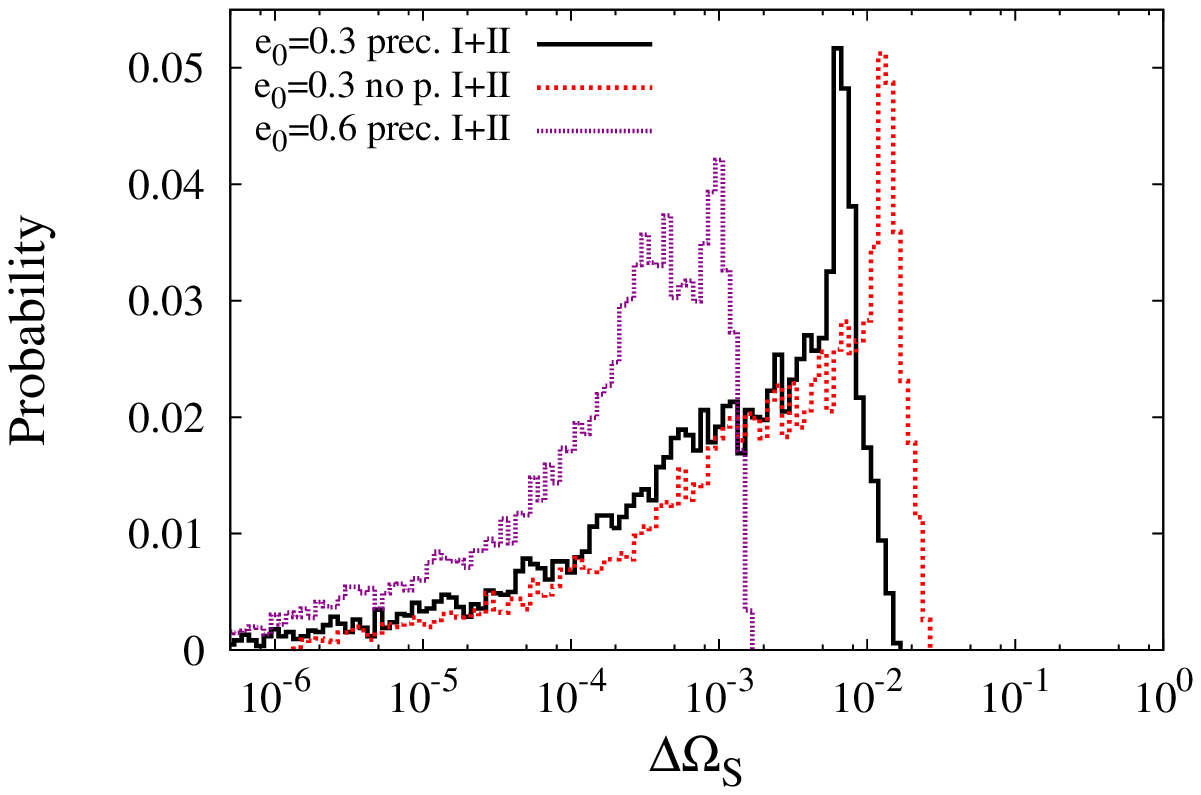}
\includegraphics[width=0.52\textwidth]{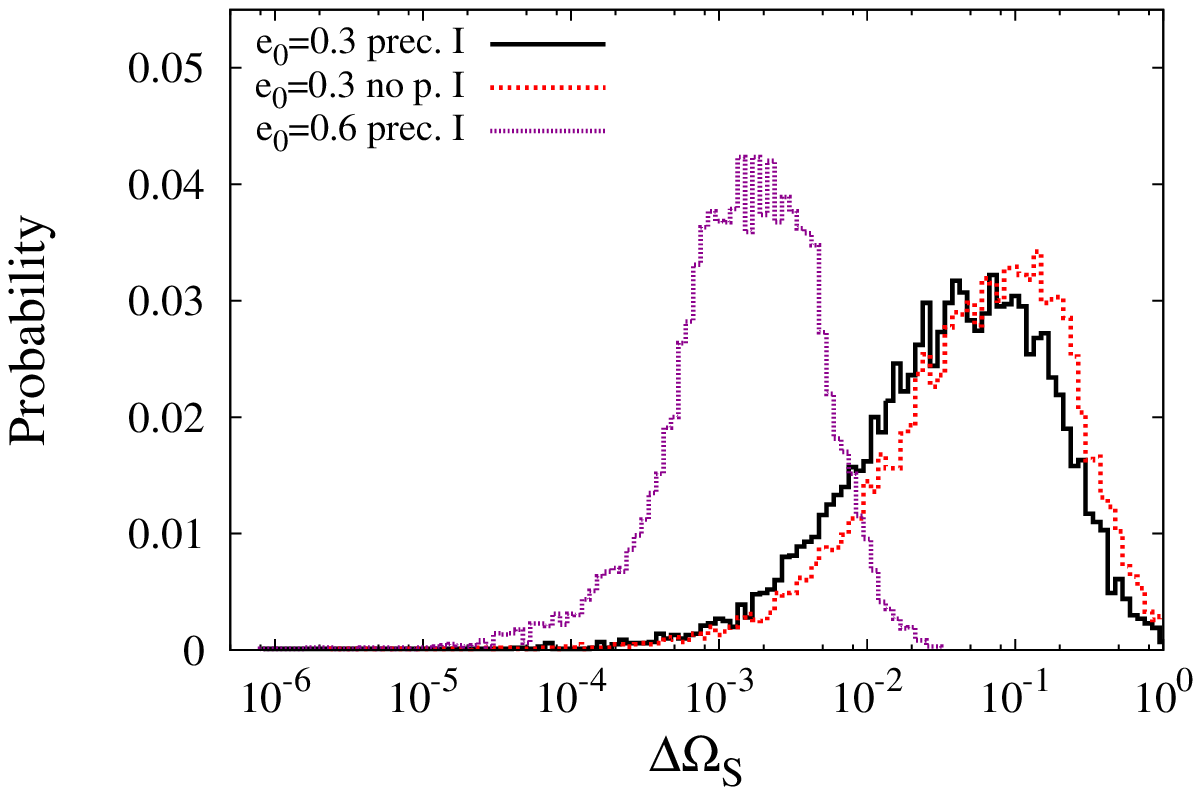}
\end{center}
\caption{(color online). Estimated distribution of the angular
resolution $\Delta\Omega_{S}$ in the precessing and nonprecessing
cases for the total ($I+II$, top) and single ($I$, bottom)
detectors. The results are shown for medium ($e_{0}=0.3$) and high
($e_{0}=0.6$) initial eccentricities and higher-mass SMBH binaries
$\left( 10^{7}-10^{7}\right) M_{\odot }$. For nonprecessing sources
the $e_{0}=0.6$ case is omitted in both figures since it is close
and very similar to the curve for the $e_{0}=0.6$ precessing case.}
\label{angular77}
\end{figure}

We have found that the LISA parameter estimation accuracy depends
sensitively on the initial eccentricity and pericenter precession,
and we have also examined the distribution of parameter errors for a
wide range of initial binary parameters and masses. The four angular
parameters
($\bar{\phi}_{S},\bar{\mu}_{S},\bar{\phi}_{L},\bar{\mu}_{L}$) are
chosen randomly in a Monte Carlo sampling, and the cosmological
redshift and luminosity distance are fixed at $z=1$ and
$D_{L}=6.4\mathrm{Gpc}$. Figures \ref{chirp66}-\ref{angular77} show
the histograms of the expected measurement errors of the binary
parameters for the chirp mass $\Delta \mathcal{M}/\mathcal{M}$,
initial eccentricity $\Delta e_{0}$, and angular resolution $\Delta
\Omega _{S}$ for equal-mass binaries with $10^{6}M_{\odot }$ or
$10^{7}M_{\odot }$ each. Our parametrization of the orbit is
singular at $e_{0}=0$. To get around this, we use $e_{0}=10^{-6} $
for circular orbits. We have presented three representative cases
for the initial eccentricity: a nearly \textit{circular} orbit with
$e_{0}=10^{-6}$ (see Table \ref{table2} and Fig. \ref{Fig3b}), and
orbits with \textit{medium} $e_{0}=0.3$ and \textit{high}
$e_{0}=0.6$ eccentricities. Our computations correspond to a 1 yr
inspiral before \textit{LSO}. The initial and final orbital
frequencies ($\nu _{0}$ and $\nu _{LSO}$) vary for the three kinds
of initial eccentricities and different equal-mass SMBH binaries as
shown in Table \ref{table1}. If the initial eccentricity $e_{0}$
increases, the initial frequency $\nu _{0}$ decreases 1 yr before
LSO, while the final frequency $\nu _{LSO}$ does not change
significantly, due to the fact that $e_{LSO}$ is close to zero.

Representative values are shown in Table \ref{table2} for equal-mass
SMBHs for a fixed set of angular configurations
($\bar{\phi}_{S}=4.642$, $\bar{\mu}_{S}=-0.3185$,
$\bar{\phi}_{L}=4.724$ and $\bar{\mu}_{L}=-0.3455$). The table shows
that accounting for the eccentricity in the waveform improves some
of the parameter errors such as the errors of the angular resolution
$\Delta \Omega _{S}$, the initial eccentricity $\Delta e_{0}$, and
the chirp mass $\Delta \mathcal{M}/\mathcal{M}$ for higher-mass SMBH
binaries $(10^{6}-10^{7})M_{\odot }$. For lower masses, i.e.
$(10^{4}-10^{5})M_{\odot } $, the eccentricity and precession have
no essential effects on parameter estimation. For masses of
$10^{4}M_{\odot }$, high eccentricity has no significant effect on
the parameters $\Delta\mathcal{M}/\mathcal{M}$ or $\Delta \Omega
_{S}$. However, the initial eccentricity errors ($\Delta e_{0}$) are
improved for smaller masses typically by factors of 3--10 and they
are greatly improved for larger initial eccentricities by orders of
magnitude. Similarly, the source localization angular resolution
$\Delta \Omega _{S}$ decreases with increasing eccentricity and
mass. However, pericenter precession does improve the parameter
errors for higher-mass SMBHs. It can be seen that the eccentricity,
compared to the circular orbit case, does improve the error of
luminosity distance $\Delta D_{L}/D_{L}$ , but there is no essential
change between the high and medium eccentricities with the inclusion
of pericenter precession. The error of $t_{c}$ is not affected by
the eccentricity or by pericenter precession. It is interesting to
note that there are degeneracies ($\Delta \Phi _{c},\Delta \gamma
_{0}>1$) for errors of $\Phi _{c}$ and $\gamma _{0}$ in the nearly
circular case, which can be explained by the fact that our
parametrization of the orbit is singular at $e_{0}=0$. For eccentric
orbits (medium and high initial eccentricities) this degeneracy
disappears(the errors of $t_{c}$, $\Phi _{c}$ and $\gamma _{0}$ are
not presented in Table \ref{table2}).

Figures \ref{SNRe} and \ref{SNRm} show the distribution of the SNR
for different binary orientations, for various eccentricities and
masses. The SNR is similar for equal-mass binaries with
$10^{5}M_{\odot }\leq M\leq 10^{7}M_{\odot }$, but significantly
smaller for a SMBH of $10^{4}M_{\odot }$ or less. Remarkably, the
SNR does not change significantly with the initial eccentricity,
which is consistent with previous studies for small eccentricities
\cite{Yunes}. This shows that the systematic improvement of the
parameter estimation accuracy for eccentric sources is due to the
breaking of correlations between different parameter errors instead
of an overall change in the SNR.

Figure \ref{Fig3b} shows the distribution of the major and/or minor
axes of the sky position error ellipse for the nearly circular,
medium and high initial eccentricity orbits. The shape of the error
ellipse is important in coordinating GW observations with telescopes
\cite{LH,Bence-apj08}. It can be seen that the error of the major
and/or minor axes is improved for highly eccentric binaries.


Figures \ref{chirp66} and \ref{chirp77} show that the chirp mass
errors are greatly improved for a larger initial eccentricity for
$10^{6}M_{\odot }$ and $10^{7}M_{\odot }$ equal-mass SMBH binaries
(see also Ref. \cite{TS}). Furthermore, the chirp mass measurement
errors are improved by an additional factor of 2--5 due to
pericenter precession for relatively massive $10^{7}M_{\odot }$
binaries, but not for $10^{6}M_{\odot }$ binaries. The typical chirp
mass error is about $10^{-5}$ for $10^{7}M_{\odot }$ and $10^{-4}$
for $10^{6}M_{\odot }$ binaries.

Figures \ref{e066} and \ref{e077} show that the initial eccentricity
errors are also improved for a high eccentricity, as the initial
eccentricity parameter can be measured with high accuracy; $\Delta
e_{0}$ is about $10^{-5}$ to $10^{-4}$ for $10^{7}M_{\odot }$
binariesm and about $10^{-4}$ to $10^{-3}$ for $10^{6}M_{\odot }$
binaries. Pericenter precession improves the eccentricity errors by
a factor of 10 for $10^{7}M_{\odot }$ and by a factor of 2--3 for
$10^{6}M_{\odot }$.

Figures \ref{angular66} and \ref{angular77} show that the typical
source sky localization accuracy for equal-mass binaries at $z=1$
ranges between $10^{-4}$ and $10^{-2}$ steradians. Consistent with
previous studies \cite{Arun,TS}, we find that the errors improve for
higher initial eccentricities ($e_{0}=0.6$), compared to the cases
of moderate to small initial eccentricities ($e_{0}=0.3$) for
equal-mass $10^{7}M_{\odot }$ binaries. The error $\Delta \Omega
_{S}$ in the total two-detector case is about 1 order of magnitude
better than for a single detector \cite{Cutler98}. For high initial
eccentricities, the angular resolution of the total detector case is
improved more compared to the single detector case for
$10^{7}M_{\odot }$ binaries\ (see Fig.\ref{angular77}). In contrast
to the chirp mass and the eccentricity errors, the angular
localization capabilities are not improved for eccentric equal-mass
$10^{6}M_{\odot}$ binaries but they are improved for
$10^{7}M_{\odot}$ binaries. Figures~\ref{angular66} and
\ref{angular77} clearly show that pericenter precession does not
affect the sky position error for either mass choice.

A possible explanation for the qualitatively different improvement
of the sky position and mass-eccentricity errors is that the sky
position is a slow parameter, as opposed to fast parameters like the
chirp mass and eccentricity \cite{Bence-prd}. The slow parameters
are determined by the slow orbital modulation of the signal by the
detector's motion around the Sun, while the fast parameters also
depend on the orbital phase. The correlations between the slow
parameters become large during the last week before merger when the
signal-to-noise ratio increases, which prohibits the rapid
improvement of the slow parameters' marginalized errors. Pericenter
precession does not vary the binary inclination and cannot
effectively break the correlation between slow parameters. However,
pericenter precession splits the GW frequency into a triplet for
each harmonic which can break degeneracies for the fast parameters
and efficiently improve their measurement errors.

\section{Conclusions}

We carried out an extensive study of parameter estimation for
eccentric binaries with arbitrary orbital eccentricity. We computed
the waveforms in the frequency domain by a new method optimized for
taking into account eccentricity, by changing the integration
variable for the waveforms from the orbital frequency $\nu(e)$ to
the eccentricity variable $e$ \cite{Bence}. This results in an
improvement of numerical precision as compared to standard
approaches in the frequency domain, where a Taylor series expansion
of the orbital frequency $\nu(e)$ (among others) in the eccentricity
$e$ is needed \cite{Yunes}. Our method is well suited for computing
the Fisher matrix and the signal-to-noise ratio. Our parameter space
is ten dimensional, consisting of four angles, the chirp mass, the
luminosity distance, coalescence time and phase, initial
eccentricity and pericenter position (compare Fig. \ref{geometry}).
The first eight parameters are standard for circular orbits too.

We have examined the LISA parameter estimation errors for GWs
emitted by eccentric inspiraling SMBH binaries including the effects
of pericenter precession. Based on a large set of simulated binary
waveforms, we found that there is about 1 order of magnitude
improvement compared to circular waveforms in LISA's angular
resolution for highly eccentric sources (e.g. $e_{0}=0.6$) for
relatively high SMBH masses $\sim 10^{7}M_{\odot }$. There is
however, a much smaller effect for lower-mass binaries in the range
$(10^{4}-10^{5})M_{\odot }$. This improves the prospects for
identifying the electromagnetic counterparts
\cite{Bence-apj06,Bence-apj08} of relatively high-mass eccentric
SMBH mergers with LISA. Similar conclusions have been reached in
Refs.~\cite{Arun,TS}. However, we found that pericenter precession
does not further improve the sky localization accuracy of the
source, although it may further improve the measurement errors of
mass and eccentricity parameters.

It is important to note that the angular resolution is significantly
affected by the number of detectors (see Figs. \ref{angular66} and
\ref{angular77}). However, nearly the same parameter estimation
accuracy can be obtained for the single and total detector
configurations for $(10^6-10^6)M_{\odot }$ binaries for fast
parameters \citep{Bence-prd} like the chirp mass and eccentricity
(Figs. \ref{chirp66} and \ref{e066}). The second detector
systematically reduces the errors of these parameters for higher
masses $(10^7-10^7)M_{\odot }$.

\begin{table*}[tb]
\caption{Parameter estimation errors for equal-mass SMBH binaries.
The initial eccentricities $e_{0}$ are $10^{-6}$ (nearly circular),
$0.3$, and $0.6.$, the luminosity distance is
$D_{L}=6.4\mathrm{Gpc}$ ($z=1$); and the angular parameters are
$\protect\phi _{L}=4.724$, $\protect\mu _{L}=-0.3455$ ,
$\protect\phi _{S}=4.642$, and $\protect\mu _{S}=-0.3185$.}
\label{table2}
\begin{center}
\begin{tabular}{c|c|ccccc}
\hline\hline $\underset{(M_{\odot })}{SMBH}$ & $e_{0}$/precession &
$SNR$ & $\underset{ (\times 10^{-2})}{\Delta D_{L}/D_{L}}$ &
$\underset{(\times 10^{-6})}{\Delta \mathcal{M}/\mathcal{M}}$ &
$\underset{(\times 10^{-6})}{\Delta e_{0}}$ & $\underset{(\times
10^{-6})}{\Delta \Omega }$ \\ \hline\hline $10^{7}-10^{7}$ &
\begin{tabular}{l}
\begin{tabular}{l}
$e_{0}=10^{-6}$, no prec. \\
$e_{0}=10^{-6}$, incl. prec.
\end{tabular}
\\
\begin{tabular}{l}
$e_{0}=0.3$, no prec. \\
$e_{0}=0.3$, incl. prec.
\end{tabular}
\\
\begin{tabular}{l}
$e_{0}=0.6$, no prec. \\
$e_{0}=0.6$, incl. prec.
\end{tabular}
\end{tabular}
&
\begin{tabular}{c}
\begin{tabular}{c}
$1119$ \\
$2002$%
\end{tabular}
\\
\begin{tabular}{c}
$1116$ \\
$1984$%
\end{tabular}
\\
\begin{tabular}{c}
$1146$ \\
$1984$
\end{tabular}
\end{tabular}
&
\begin{tabular}{c}
\begin{tabular}{c}
$837$ \\
$538$
\end{tabular}
\\
\begin{tabular}{c}
$96.2$ \\
$42.9$%
\end{tabular}
\\
\begin{tabular}{c}
$31.6$ \\
$17.3$
\end{tabular}
\end{tabular}
&
\begin{tabular}{c}
\begin{tabular}{c}
$105$ \\
$9.14$%
\end{tabular}
\\
\begin{tabular}{c}
$67.7$ \\
$9.42$
\end{tabular}
\\
\begin{tabular}{c}
$17.4$ \\
$4.95$
\end{tabular}
\end{tabular}
&
\begin{tabular}{c}
\begin{tabular}{c}
$1794$ \\
$1311$
\end{tabular}
\\
\begin{tabular}{c}
$222$ \\
$34.7$
\end{tabular}
\\
\begin{tabular}{c}
$6.91$ \\
$2.14$
\end{tabular}
\end{tabular}
&
\begin{tabular}{c}
\begin{tabular}{c}
$193$ \\
$77.9$
\end{tabular}
\\
\begin{tabular}{c}
$3.32$ \\
$0.893$%
\end{tabular}
\\
\begin{tabular}{c}
$2.16$ \\
$0.689$
\end{tabular}
\end{tabular}
\\ \hline
$10^{6}-10^{6}$ &
\begin{tabular}{l}
\begin{tabular}{l}
$e_{0}=10^{-6}$, no prec. \\
$e_{0}=10^{-6}$, incl. prec.
\end{tabular}
\\
\begin{tabular}{l}
$e_{0}=0.3$, no prec. \\
$e_{0}=0.3$, incl. prec.
\end{tabular}
\\
\begin{tabular}{l}
$e_{0}=0.6$, no prec. \\
$e_{0}=0.6$, incl. prec.
\end{tabular}
\end{tabular}
&
\begin{tabular}{c}
\begin{tabular}{c}
$1171$ \\
$1704$
\end{tabular}
\\
\begin{tabular}{c}
$1176$ \\
$1701$
\end{tabular}
\\
\begin{tabular}{c}
$1200$ \\
$1712$
\end{tabular}
\end{tabular}
&
\begin{tabular}{c}
\begin{tabular}{c}
$192$ \\
$168$
\end{tabular}
\\
\begin{tabular}{c}
$30.6$ \\
$26.0$%
\end{tabular}
\\
\begin{tabular}{c}
$10.3$ \\
$8.29$
\end{tabular}
\end{tabular}
&
\begin{tabular}{c}
\begin{tabular}{c}
$3.09$ \\
$1.19$
\end{tabular}
\\
\begin{tabular}{c}
$3.99$ \\
$1.51$
\end{tabular}
\\
\begin{tabular}{c}
$3.17$ \\
$1.56$
\end{tabular}
\end{tabular}
&
\begin{tabular}{c}
\begin{tabular}{c}
$1562$ \\
$1363$
\end{tabular}
\\
\begin{tabular}{c}
$7.53$ \\
$3.32$
\end{tabular}
\\
\begin{tabular}{c}
$1.18$ \\
$0.917$
\end{tabular}
\end{tabular}
&
\begin{tabular}{c}
\begin{tabular}{c}
$13.5$ \\
$9.33$
\end{tabular}
\\
\begin{tabular}{c}
$2.00$ \\
$1.00$
\end{tabular}
\\
\begin{tabular}{c}
$1.84$ \\
$0.871$
\end{tabular}
\end{tabular}
\\ \hline
$10^{5}-10^{5}$ &
\begin{tabular}{l}
\begin{tabular}{l}
$e_{0}=10^{-6}$, no prec. \\
$e_{0}=10^{-6}$, incl. prec.
\end{tabular}
\\
\begin{tabular}{l}
$e_{0}=0.3$, no prec. \\
$e_{0}=0.3$, incl. prec.
\end{tabular}
\\
\begin{tabular}{l}
$e_{0}=0.6$, no prec. \\
$e_{0}=0.6$, incl. prec.
\end{tabular}
\end{tabular}
&
\begin{tabular}{c}
\begin{tabular}{c}
$1924$ \\
$2183$
\end{tabular}
\\
\begin{tabular}{c}
$1925$ \\
$2184$
\end{tabular}
\\
\begin{tabular}{c}
$1920$ \\
$2188$
\end{tabular}
\end{tabular}
&
\begin{tabular}{c}
\begin{tabular}{c}
$314$ \\
$296$
\end{tabular}
\\
\begin{tabular}{c}
$33.4$ \\
$26.6$
\end{tabular}
\\
\begin{tabular}{c}
$14.3$ \\
$12.0$
\end{tabular}
\end{tabular}
&
\begin{tabular}{c}
\begin{tabular}{c}
$1.03$ \\
$0.958$
\end{tabular}
\\
\begin{tabular}{c}
$1.30$ \\
$1.16$
\end{tabular}
\\
\begin{tabular}{c}
$1.04$ \\
$1.23$
\end{tabular}
\end{tabular}
&
\begin{tabular}{c}
\begin{tabular}{c}
$2595$ \\
$2365$
\end{tabular}
\\
\begin{tabular}{c}
$2.74$ \\
$3.54$
\end{tabular}
\\
\begin{tabular}{c}
$0.435$ \\
$0.831$
\end{tabular}
\end{tabular}
&
\begin{tabular}{c}
\begin{tabular}{c}
$30.6$ \\
$25.6$
\end{tabular}
\\
\begin{tabular}{c}
$0.848$ \\
$0.553$
\end{tabular}
\\
\begin{tabular}{c}
$0.678$ \\
$0.520$
\end{tabular}
\end{tabular}
\\ \hline
$10^{4}-10^{4}$ &
\begin{tabular}{l}
\begin{tabular}{l}
$e_{0}=10^{-6}$, no prec. \\
$e_{0}=10^{-6}$, incl. prec.
\end{tabular}
\\
\begin{tabular}{l}
$e_{0}=0.3$, no prec. \\
$e_{0}=0.3$, incl. prec.
\end{tabular}
\\
\begin{tabular}{l}
$e_{0}=0.6$, no prec. \\
$e_{0}=0.6$, incl. prec.
\end{tabular}
\end{tabular}
&
\begin{tabular}{c}
\begin{tabular}{c}
$306$ \\
$314$
\end{tabular}
\\
\begin{tabular}{c}
$310$ \\
$318$
\end{tabular}
\\
\begin{tabular}{c}
$333$ \\
$341$
\end{tabular}
\end{tabular}
&
\begin{tabular}{c}
\begin{tabular}{c}
$746$ \\
$697$
\end{tabular}
\\
\begin{tabular}{c}
$71.2$ \\
$62.9$
\end{tabular}
\\
\begin{tabular}{c}
$30.0$ \\
$27.3$
\end{tabular}
\end{tabular}
&
\begin{tabular}{c}
\begin{tabular}{c}
$0.628$ \\
$1.93$
\end{tabular}
\\
\begin{tabular}{c}
$0.847$ \\
$1.80$
\end{tabular}
\\
\begin{tabular}{c}
$0.539$ \\
$0.925$
\end{tabular}
\end{tabular}
&
\begin{tabular}{c}
\begin{tabular}{c}
$4605$ \\
$4433$
\end{tabular}
\\
\begin{tabular}{c}
$1.60$ \\
$4.68$
\end{tabular}
\\
\begin{tabular}{c}
$0.193$ \\
$0.356$
\end{tabular}
\end{tabular}
&
\begin{tabular}{c}
\begin{tabular}{c}
$239$ \\
$189$
\end{tabular}
\\
\begin{tabular}{c}
$30.3$ \\
$29.0$
\end{tabular}
\\
\begin{tabular}{c}
$28.3$ \\
$27.4$
\end{tabular}
\end{tabular}
\\ \hline\hline
&  &  &  &  &  &
\end{tabular}
\end{center}
\end{table*}

\acknowledgments We thank L\'aszl\'o Gond\'an for carefully reading
the manuscript and for useful discussions. This work was supported
by Hungarian Scientific Research Fund (OTKA) Grants No. NI68228 and
No. K101709. B.K. acknowledges support from NASA through Einstein
Postdoctoral Fellowship Grant No. PF9-00063 issued by the Chandra
X-ray Observatory Center, which is operated by the Smithsonian
Astrophysical Observatory for and on behalf of the National
Aeronautics Space Administration under Contract No. NAS8-03060.

\appendix

\section{Orbital evolution and waveform}

\label{appendixA}

\begin{figure}[tbh]
\begin{center}
\includegraphics[width=0.42\textwidth]{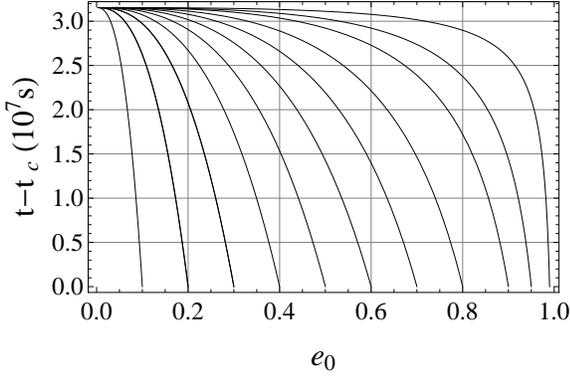}
\end{center}
\caption{The evolution of the eccentricity as a function of time (as
"lifetime" for the fixed 1 yr inspiraling time). The eccentricity
changes significantly near the coalescence.} \label{life}
\end{figure}

\begin{figure}[tbh]
\begin{center}
\includegraphics[width=0.42\textwidth]{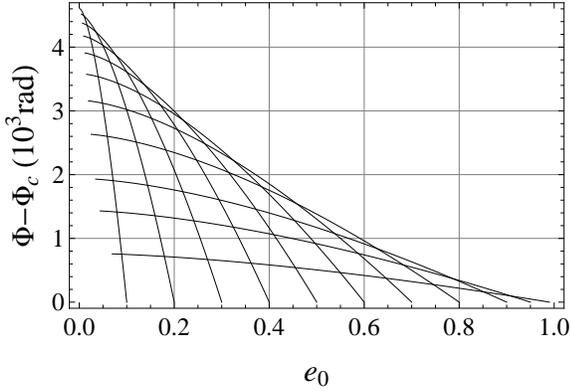}
\end{center}
\caption{The evolution of the eccentricity in terms of the phase
function for the fixed 1 yr inspiraling time.} \label{phase}
\end{figure}

$\allowbreak \allowbreak $ According to Eqs. (\ref{PP1},\ref{PP2}),
the equation
\begin{equation}
\frac{d\nu }{de}=-\frac{18\nu }{19}\frac{1+\frac{73}{24}e^{2}+\frac{37}{96}
e^{4}}{e(1-e^{2})\left( 1+\tfrac{121}{304}e^{2}\right) }\ ,  \label{dnu}
\end{equation}
can be solved as
\begin{equation}
\nu (e)=C_{0}e^{-18/19}\left( 1-e^{2}\right) ^{3/2}\left( 1+\frac{121}{304}
e^{2}\right) ^{-1305/2299}\ ,  \label{sol}
\end{equation}
where $C_{0}=\nu _{0}e_{0}^{18/19}\left( 1+\frac{121}{304}e_{0}^{2}\right)
^{1305/2299}\left( 1-e_{0}^{2}\right) ^{-3/2}$ is the integration constant
that has been chosen to set the initial condition $\nu (e_{0})=\nu _{0}$ for
the initial values $e_{0}$ and $\nu _{0}$. Then Eq. (\ref{sol}) is
\begin{equation}
\nu (e)=\nu _{0}\frac{\sigma (e)}{\sigma (e_{0})}\ ,  \label{sol2}
\end{equation}
where $\sigma (e)=e^{-18/19}\left( 1-e^{2}\right) ^{3/2}\left(
1+\frac{121}{ 304}e^{2}\right) ^{-1305/2299}$. From Eqs.
(\ref{PP1},\ref{PP2}) one can compute the evolution of the time and
phase functions [$t-t_{c}=\int_{0}^{e}\frac{de^{\prime
}}{\dot{e}(e^{\prime })}$, $\Phi -\Phi _{c}=2\pi
\int_{0}^{e}\frac{\nu (e^{\prime })}{\dot{e}(e^{\prime })}de$] in
terms of eccentricity as [see Eqs. (\ref{PP2},\ref{sol2})]
\begin{eqnarray}
t-t_{c} &=&-\frac{15}{304\mathcal{M}^{5/3}}\left( \frac{\sigma (e_{0})}{2\pi
\nu _{0}}\right) ^{8/3}I_{t}(e)  \label{egzakt1} \\
\Phi -\Phi _{c} &=&-\frac{15}{304\mathcal{M}^{5/3}}\left( \frac{\sigma
(e_{0})}{2\pi \nu _{0}}\right) ^{5/3}I_{\phi }(e)\ ,  \label{egzakt2}
\end{eqnarray}

where the $I_{t}$ and $I_{\phi }$ integrals are
\begin{eqnarray}
I_{t}(e) &=&\overset{e}{\underset{0}{\int }}\frac{x^{\alpha }\left( 1-\delta
x^{2}\right) ^{-\beta }}{(1-x^{2})^{3/2}}dx\ ,  \label{it} \\
I_{\phi }(e) &=&\overset{e}{\underset{0}{\int
}}\frac{x^{\widetilde{\alpha }} }{\left( 1-\delta x^{2}\right)
^{\widetilde{\beta }}}dx\ ,  \label{iphi}
\end{eqnarray}
with the constants $\alpha =29/19$, $\beta =-1181/2299$, $\delta
=-121/304$, $\widetilde{\alpha }=11/19$ and $\widetilde{\beta
}=124/2299$. The integrals in Eqs. (\ref{it},\ref{iphi}) can be
evaluated with the Appell functions which generalize the
hypergeometric functions \cite{Pierro2,Appell}
\begin{eqnarray}
I_{t}(e) &=&\frac{19e^{48/19}}{48}F_{1}\left( \frac{\alpha +1}{2},\beta ,
\frac{3}{2},\frac{\alpha +3}{2};\delta e^{2},e^{2}\right) \ , \\
I_{\phi }(e) &=&\frac{19e^{30/19}}{30}\,_{2}F_{1}\left( \frac{\widetilde{
\alpha }+1}{2},\widetilde{\beta },\frac{\widetilde{\alpha }+3}{2};\delta
e^{2}\right) \ .
\end{eqnarray}
To compute the time ($\Delta T$) and phase ($\Delta \Phi $)
differences the binary spends between the initial and final
eccentricities $e_{0}$ and $e_{1} $ during its evolution, Eqs.
(\ref{egzakt1},\ref{egzakt2}) are used,
\begin{eqnarray}
\Delta T &=&\frac{15}{304\mathcal{M}^{5/3}}\left( \frac{\sigma (e_{0})}{2\pi
\nu _{0}}\right) ^{8/3}\left[ I_{t}(e_{0})-I_{t}(e_{1})\right] \ ,
\label{ker1} \\
\Delta \Phi &=&\frac{15}{304\mathcal{M}^{5/3}}\left( \frac{\sigma (e_{0})}{
2\pi \nu _{0}}\right) ^{5/3}\left[ I_{\phi }(e_{0})-I_{\phi }(e_{1})\right]
\ .  \label{ker2}
\end{eqnarray}
Figures \ref{life} and \ref{phase} show the evolution of time and
phase for various initial eccentricities, a fixed 1 yr inspiraling
time before the \textit{LSO} and $10^{6}M_{\odot }$ equal-mass
binaries. It can be seen that the eccentricity changes significantly
near the coalescence, and the accumulated number of orbits is
decreasing for high initial eccentricity.

\section{Stationary Phase Approximation}

\label{appendixB}

Consider the waveform $h(t)=\mathcal{A}(t)\cos \Phi (t)$ with
$\mathcal{\dot{A}}(t)/\mathcal{A}(t)\ll \dot{\Phi}(t)$ and
$\ddot{\Phi}(t)\ll \dot{\Phi}(t)^{2}$ (see e.g. Ref. \cite{FC}),
with its Fourier transform as
\begin{equation}
\mathcal{F}\left[ \mathcal{A}(t)\cos \Phi (t)\right]
=\underset{-\infty }{\overset{\infty }{\int
}}\mathcal{A}(t)\frac{e^{i\Phi (t)}+e^{-i\Phi (t)}}{2}e^{2\pi
itf}dt\ .  \label{fourier}
\end{equation}

To evaluate the Fourier integral one can use the \textit{stationary
phase approximation} (SPA). For an arbitrary function of the time,
$\Psi (t)$, $\underset{-\infty }{\overset{\infty }{\int
}}\mathcal{A}(t)e^{i\Psi (t)}dt\simeq
\mathcal{A}(\mathcal{T})\sqrt{2\pi
/\ddot{\Psi}(\mathcal{T})}e^{i\left( \Psi
(\mathcal{T})+\mathrm{sign}\left[ \ddot{\Psi}(\mathcal{T})\right]
\pi /4\right) }$, where the saddle point $\mathcal{T} $ satisfies
$\dot{\Psi}(\mathcal{T})=0$. In Eq. (\ref{fourier}), the $e^{i\Phi
(t)}$ terms have no contributions to the saddle point $\mathcal{T}$
. Moreover, $\Psi (t)=2\pi tf-\Phi (t)$, and the stationary phase
condition [$\dot{\Psi}(\mathcal{T})=0$] implies that
$f=\dot{\Phi}(\mathcal{T})/(2\pi )$. This provides a relation
between the Fourier and orbital frequencies. Carrying out this
exercise for an eccentric waveform consisting of many widely
separated GW harmonics, the corresponding Fourier frequencies are,
respectively, $f_{n}=n\nu $ and $f_{n\pm }=n\nu \pm \dot{\gamma}/\pi
$ for the terms due to pericenter precession. For circular orbits,
the only nonvanishing term has frequency $f=2\nu $. Therefore, the
Fourier transform of harmonic functions with SPA are
\begin{eqnarray}
\mathcal{F}\left[ \mathcal{A}(t)\sin \Phi (t)\right]
&=&\tfrac{\mathcal{A} \left[ f(\mathcal{T})\right]
}{2}\sqrt{\tfrac{2\pi }{\left\vert \ddot{\Psi} \left[
f(\mathcal{T})\right] \right\vert }}e^{i\left( \Psi \left[
f(\mathcal{\ T})\right] +\frac{\pi }{4}\right) }\ ,  \label{int1} \\
\mathcal{F}\left[ \mathcal{A}(t)\cos \Phi (t)\right]
&=&\tfrac{\mathcal{A}\left[ f(\mathcal{T})\right]
}{2}\sqrt{\tfrac{2\pi }{\left\vert \ddot{\Psi} \left[
f(\mathcal{T})\right] \right\vert }}e^{i\left( \Psi \left[
f(\mathcal{\ T})\right] -\frac{\pi }{4}\right) }\ ,  \label{int2}
\end{eqnarray}
where $\Psi \left[ f(\mathcal{T})\right] =2\pi f(\mathcal{T})t\left[
\nu (\mathcal{T})\right] -\Phi \left[ \nu (\mathcal{T})\right] $ is
the phase function and $t\left[ \nu (\mathcal{T})\right] $, $\Phi
\left[ \nu (\mathcal{T})\right] $ are derived from radiation
reaction by Eqs. (\ref{egzakt1},\ref{egzakt2}).

Following Ref. \cite{ecc2}, the phase functions for eccentric
compact binaries are
\begin{eqnarray}
\Psi _{n} &=&2\pi ft-\Phi _{n}\ ,\text{ } \\
\Psi _{n\pm } &=&2\pi ft-\Phi _{n\pm }\ ,
\end{eqnarray}
where the functions $\Phi _{n},\Phi _{n\pm }$ are defined by Eqs.
(\ref{phase1},\ref{phase2}) and the first time derivatives are
expressed as
\begin{eqnarray}
\dot{\Psi}_{n} &=&2\pi f-2\pi n\nu \ ,\text{ }  \label{s1} \\
\dot{\Psi}_{n\pm } &=&2\pi f-2\pi n\nu \mp 2\dot{\gamma}\ .  \label{s2}
\end{eqnarray}
There are three saddle points ($t_{n}$, $t_{n\pm }$) following from
the stationary phase conditions $\dot{\Psi}_{n}(t_{n})=0$ and
$\dot{\Psi}_{n\pm }(t_{n\pm })=0$. It follows that there are three
Fourier frequencies for each harmonic of the orbital frequency
(denoted by $f_{n}$, $f_{n\pm }$). The second time derivatives of
the $\Psi _{n}$ and $\Psi _{n\pm }$ phase functions are
\begin{eqnarray}
\ddot{\Psi}_{n} &=&-2\pi n\dot{\nu}\ ,\text{ } \\
\ddot{\Psi}_{n\pm } &=&-2\pi n\dot{\nu}\mp 2\ddot{\gamma}\ ,
\end{eqnarray}
where $\ddot{\gamma}$ is the time derivative of $\dot{\gamma}$
induced by gravitational radiation; see Eqs. (\ref{PP1},\ref{PP2}).
Then the phase functions of the waveforms, Eqs.
(\ref{elll1},\ref{elll2}), can be expressed in terms of the time
corresponding to the stationary phase and the acceleration of the
pericenter precession, formally
\begin{eqnarray}
\Psi _{n}(f_{n}) &=&2\pi f_{n}t_{n}(f_{n})-\Phi _{n}(f_{n})\ , \\
\Psi _{n\pm }(f_{n\pm }) &=&2\pi f_{n\pm }t_{n\pm }(f_{n\pm })-\Phi _{n\pm
}(f_{n\pm })\ .
\end{eqnarray}

\end{document}